\documentclass[journal]{IEEEtran}
\usepackage{amsmath,amsfonts}

\usepackage{algorithm}
\usepackage{algpseudocode}
\usepackage{array}
\usepackage[caption=false,font=normalsize,labelfont=sf,textfont=sf]{subfig}
\usepackage{textcomp}
\usepackage{stfloats}
\usepackage{url}
\usepackage{verbatim}
\usepackage{graphicx}
\usepackage{cite}
\usepackage{tikz}
\usepackage{booktabs}
\usepackage{multirow}
\usepackage{float}
\usepackage{amssymb}

\allowdisplaybreaks[4]
\begin{document}

\title{Embodied Intelligence for Spectrum Map Construction in Low-Altitude Wireless Networks}

\author{\IEEEauthorblockN{Hongtao Liang, \emph{Graduate Student Member}, \emph{IEEE}, Yiyao Wan, Jiahuan Ji, \emph{Member}, \emph{IEEE}, \\
Lingda Meng, Fuhui Zhou, \emph{Senior Member}, \emph{IEEE}, and
Qihui Wu, \emph{Fellow}, \emph{IEEE}}

\thanks{H. Liang, L. Meng and Q. Wu are with the College of Electronic and Information Engineering, Nanjing University of Aeronautics and Astronautics, Nanjing, 210000, P. R. China. (e-mail: ceie.lht@nuaa.edu.cn, 042300229@nuaa.edu.cn, wuqihui2014@sina.com).

Y. Wan, F. Zhou and J. Ji are with the College of Artificial Intelligence, Nanjing University of Aeronautics and Astronautics, Nanjing, 210000, P. R. China. (e-mail: yiyaowan@nuaa.edu.cn, zhoufuhui@ieee.org, jiahuan.ji@nuaa.edu.cn).}
}

\maketitle
\begin{abstract}
\bfseries
Accurate spectrum map construction by unmanned aerial vehicles (UAVs) is crucial for intelligent resource management in low-altitude wireless networks (LAWNs).
Although various construction approaches have been investigated, their performance often degrades under extremely sparse sampling because purely data-driven models may not capture the underlying propagation physics.
To address these challenges, an embodied intelligence framework is proposed, which synergistically integrates closed-loop perception-cognition-execution (PCE) coupling and embodied exploration to achieve precise and efficient spectrum construction.
Based on this framework, a corresponding embodied cognition and decision network (ECDNet) with three tightly coupled modules is designed. Specifically, by leveraging distance-aware convolutions, the semantic bridge, and adaptive feature alignment mechanisms, ECDNet establishes the bidirectional information flow between construction and decision-making. 
Simulation results show that ECDNet achieves a root mean square error (RMSE) of 2.787~dBm at 4.00\% sampling under the non-revisiting spiral protocol. 
Furthermore, a practical UAV-based spectrum construction platform is developed and field-deployed  to provide prototype-level evidence that the sensing, cognition, and action modules execute as a closed loop on size, weight, and power constrained UAV edge devices. 

\end{abstract}


\section{Introduction}

\IEEEPARstart{L}{ow-altitude} wireless networks (LAWNs) are emerging as critical infrastructure for next-generation intelligent systems~\cite{BaiCOMST2023,FanIoTJ2025,Zeng2016,NguyenJSTSP2022}. Despite offering significant application benefits, the proliferation of dense unmanned aerial vehicles (UAVs) creates unprecedented spectrum management challenges. To enable intelligent resource allocation and interference mitigation in such complex LAWNs, accurate spectrum map construction using UAVs is essential to provide comprehensive spatial-frequency profiles of the electromagnetic environment~\cite{ShresthaTSP2022}. However, UAV platforms are subject to stringent size, weight, and power constraints that severely limit operational endurance. These constraints restrict UAVs to collecting measurements at only a small fraction of spatial locations, typically below 5\% sampling density relative to the target map resolution~\cite{TimilsinaTSP2024}. Consequently, constructing high-fidelity spectrum maps from such sparse measurements is of paramount importance.

Traditional approaches for spectrum map construction predominantly adopt a purely data-driven decoupled paradigm~\cite{LiangArXiv2025Synergy}. This paradigm separates spectrum map construction into two independent modules, namely, the construction module for inferring the continuous spectrum map from limited data~\cite{ref_1,RomeroSPM2022,ref_3,ACTGAN_IET_2024,STORM_Trans_2024,LuICML2024,RaissiJCP2019,ShahidDySPAN2025,ref_2,ref_4,ZengWC2021_CKM} and the sampling module for selecting informative sensing locations~\cite{ZhangTWC2021,DongSPL2022,PolyzosTWC2024,ShresthaTWC2023,ZengTWC2021,EsrafilianGLOBECOM2021}.  Although this modularity simplifies the independent implementation of construction and sampling, it can limit joint optimization when no mechanism is provided to exchange the evolving construction state with the sampling policy. Specifically, in approaches without online construction-to-sampling feedback, the current construction state does not directly guide the next sensing decision~\cite{ZhangTWC2021,DongSPL2022}. Moreover, data-driven construction methods may struggle to produce physically plausible results under extreme data scarcity, generating unrealistic artifacts that do not reflect the underlying propagation characteristics~\cite{ref_3,ACTGAN_IET_2024,STORM_Trans_2024}.

Embodied intelligence (EI) provides a promising framework to address these challenges by establishing the closed-loop coupling between perception, cognition, and execution (PCE)~\cite{LiangArXiv2025Synergy,LiuCSUR2025}. This mechanism transforms the passive observation into the active exploration, where cognitive insights from the construction process can directly inform sampling decisions.  This enables the UAV agent to dynamically adjust its sampling decisions based on the evolving construction state and collect additional measurements to reduce construction errors, thereby supporting more efficient use of the limited sampling budget. Moreover, physics-informed methods incorporate wireless propagation knowledge into the construction process through architectural or regularization priors, which can promote physically plausible spectrum-map estimates under sparse measurements~\cite{RaissiJCP2019,ShahidDySPAN2025}.

Motivated by these advantages, applying EI to UAV spectrum map construction is a promising direction for breaking the decoupled paradigm and mitigating data scarcity. However, adapting this unified approach for spectrum construction faces specific challenges due to the unique characteristics of LAWNs. Specifically, different from the traditional vision-based EI system that receives dense pixel-level feedback, the spectrum construction relies on sparse scalar received signal strength indication (RSSI) values collected from discontinuous spatial locations, which hinders the effective PCE coupling.  Moreover, purely data-driven methods may struggle to capture complex electromagnetic propagation patterns from extremely sparse measurements. Consequently, data scarcity can increase the risk of physically implausible spectrum-map estimates and poor generalization~\cite{ShresthaTSP2022}.  Furthermore, complex LAWN environments are susceptible to high levels of noise and interference. The resulting low signal-to-noise ratio (SNR) further exacerbates the construction challenge, necessitating a robust method capable of recovering reliable signals from sparse and noisy measurements~\cite{ShaoTMC2024Jamming}.

To address these challenges, we propose an EI-based framework for UAV spectrum map construction. The framework is realized through a corresponding embodied cognition and decision network (ECDNet). The main contributions of this paper are summarized as follows.

\begin{itemize}
\item An EI-based framework is proposed for spectrum map construction.
Specifically, a synergistic closed-loop PCE coupling is established as the core mechanism to bridge the information gap between construction and decision-making.
Moreover, physics-informed priors are integrated as a supporting technique to enhance generalization under sparse measurements.
 Based on the free energy principle (FEP), this framework provides a unified optimization perspective for jointly organizing the construction process and the exploration strategy through task-specific construction and decision objectives.

\item An ECDNet is designed to effectively implement the unified framework through three tightly coupled components.
Specifically, the environment perception component fuses multi-modal sensor measurements into structured inputs.
 Moreover, the situational cognition component encodes propagation-related structure by using distance-aware convolutions to construct the spectrum map from sparse data. 
Furthermore, the active exploration component utilizes a semantic bridge to extract cognitive insights for intelligent sampling decisions.
These components establish a closed PCE loop, enabling the UAV to autonomously achieve spectrum map construction under resource-constrained conditions.

\item A practical UAV-based spectrum construction platform is developed to support evaluation across controlled simulations and prototype-level field demonstrations.
Specifically, simulation results demonstrate superior sample efficiency under the non-revisiting spiral protocol, achieving an RMSE of 2.787~dBm at a sparse 4.00\% sampling rate.
 Furthermore, the field tests provide prototype-level evidence of real-time closed-loop operation on size, weight, and power (SWaP) constrained UAV edge devices. 

\end{itemize}

The remainder of this paper is organized as follows. Section II reviews related work. Section III presents the EI-based spectrum map construction framework. Section IV details the ECDNet architecture. Section V presents experimental results and field deployment. Section VI concludes the paper.

\section{Related Work}

To enable precise spectrum map construction, extensive research efforts have been conducted by using various construction strategies. Specifically, current methods can be categorized into static observation-based methods \cite{ref_1,RomeroSPM2022,ref_3,ACTGAN_IET_2024,STORM_Trans_2024,LuICML2024,RaissiJCP2019,ShahidDySPAN2025,ref_2,ref_4,ZengWC2021_CKM} and dynamic exploration-based methods \cite{ZhangTWC2021,DongSPL2022,PolyzosTWC2024,ShresthaTWC2023,ZengTWC2021,EsrafilianGLOBECOM2021}.

\subsection{Static Observation-Based Construction Methods}

Static observation-based methods deploy sensor networks or predetermined measurement campaigns and focus on construction algorithms that infer continuous fields from distributed stationary observations.  These approaches include classical geostatistical techniques \cite{ref_1,RomeroSPM2022}, deep learning methods \cite{ref_3,ACTGAN_IET_2024,STORM_Trans_2024}, and frameworks incorporating physical or semantic information \cite{RaissiJCP2019,ShahidDySPAN2025,ref_2,ref_4,ZengWC2021_CKM,LuICML2024}.  In particular, early geostatistical approaches treat the spectrum map estimation as spatial regression under stationarity assumptions\cite{ref_1,RomeroSPM2022}. The authors in \cite{ref_1} proposed space-frequency-interpolated radio maps to jointly exploit spatial and cross-frequency correlations, providing principled uncertainty quantification.  However, their accuracy can degrade when stationarity assumptions poorly approximate heterogeneous urban propagation.

 Standard batch kriging may incur cubic computational cost in the number of observations. Its estimates can support active measurements when combined with an acquisition or planning mechanism.

To overcome these limitations, data-driven deep learning approaches have been extensively explored \cite{ref_3,ACTGAN_IET_2024,STORM_Trans_2024}. Specifically, encoder-decoder architectures with skip connections were developed in \cite{ref_3} to enable end-to-end learning but yield over-smoothed estimations that lack fine-grained spatial details. Building upon this architecture, the aggregated contextual-transformation generative adversarial network was proposed in \cite{ACTGAN_IET_2024} to improve the perceptual quality through context-aware attention. The authors in \cite{STORM_Trans_2024} introduced spatial transformers to enable gridless interpolation with robustness to spatial transformations.  Nevertheless, extremely sparse measurements can reduce construction accuracy and increase the risk of physically implausible estimates.

Physics-informed and semantic-enhanced approaches have emerged to address this fundamental limitation \cite{LuICML2024,RaissiJCP2019,ShahidDySPAN2025,ref_2,ref_4,ZengWC2021_CKM}. Physics-informed neural networks were introduced in \cite{RaissiJCP2019} to embed partial differential equation residuals as soft constraints in the loss function. Extending this to wireless scenarios, the authors in \cite{ShahidDySPAN2025} incorporated RSSI partial differential equations derived from path-loss models. Complementing physical priors, frequency-spatial reasoning was developed in \cite{ref_2} to infer missing bands by exploiting cross-frequency correlations, while the authors in \cite{ref_4} employed multi-task learning to jointly optimize map construction and source localization. The channel knowledge map concept was introduced in \cite{ZengWC2021_CKM} as a unifying framework for environment-aware representations. Despite the progressive sophistication, all static methods share a fundamental limitation where construction quality critically depends on the spatial distribution of sampling locations \cite{RomeroSPM2022}.  When used as standalone construction modules, these methods do not specify subsequent measurement locations and require an accompanying strategy for active acquisition.   Collecting dense measurements across a target region can be costly, motivating adaptive sampling strategies \cite{PolyzosTWC2024}.

\subsection{Dynamic Exploration-Based Construction Methods}

Dynamic exploration-based methods transform passive measurement into active data acquisition, where UAVs dynamically optimize sampling trajectories to maximize information gain rather than following fixed paths. Research has evolved from map-assisted planning \cite{ZhangTWC2021,DongSPL2022} and uncertainty-driven active learning \cite{PolyzosTWC2024,ShresthaTWC2023} to deep reinforcement learning (DRL) for joint optimization \cite{ZengTWC2021,EsrafilianGLOBECOM2021}, with recent perspectives \cite{LiangArXiv2025Synergy} emphasizing tight PCE coupling.

Map-assisted approaches utilize precomputed spectrum maps for spectrum-aware trajectory design to optimize communication objectives \cite{ZhangTWC2021,DongSPL2022}. Specifically, \cite{ZhangTWC2021} formulated three-dimensional path planning to guarantee connectivity for cellular-connected UAVs. Similarly, radio maps were exploited in \cite{DongSPL2022} to design anti-jamming trajectories balancing quality-of-service and resilience. While demonstrating the value of spectrum information, these methods rely on static maps without online updates. Consequently, they exhibit unidirectional information flow, representing a decoupled paradigm that fails to support active map construction.

Uncertainty-driven active learning methods have been developed to address the interdependence between map quality and sampling strategy \cite{PolyzosTWC2024,ShresthaTWC2023}. The Bayesian active learning was proposed in \cite{PolyzosTWC2024} to design information-gain acquisition functions over Gaussian process models. Extending this framework to aerial platforms, the authors in \cite{ShresthaTWC2023} formulated spectrum surveying with UAVs to co-design paths and sampling through information-theoretic objectives. These methods achieve significant measurement reduction. However, construction and decision modules remain separated, with uncertainty provided as an explicit estimator output rather than extracted from internal network representations.

DRL methods have emerged to achieve tighter integration through end-to-end policy optimization \cite{ZengTWC2021,EsrafilianGLOBECOM2021}. The authors in \cite{ZengTWC2021} investigated simultaneous navigation and radio mapping where the agent progressively builds maps while learning motion and measurement strategies. To improve sample efficiency, model-aided approaches were proposed in \cite{EsrafilianGLOBECOM2021} to incorporate approximate physical models. These methods represent the closest paradigm to our work. However, they suffer from shallow coupling where only low-dimensional geometric features serve as agent state and interaction occurs through scalar reward signals. Compared with robotics active SLAM~\cite{AhmedSensors2023} that achieves dense pixel-level feedback, spectrum construction faces unique challenges where sparse scalar RSSI measurements create a severe information bottleneck and electromagnetic propagation physics demands domain-specific priors. Recent perspectives emphasize that tight PCE coupling is the hallmark of embodied systems \cite{LiuCSUR2025}. Our work addresses these gaps by proposing an EI framework with closed-loop PCE coupling that organizes spectrum construction and active exploration under an FEP-motivated optimization perspective through the semantic bridge.

\section{EI-Based UAV Spectrum Map Construction Framework}

In this section, the EI-based UAV spectrum construction framework is presented. Then, we provide the mathematical formulation for the joint PCE optimization.

\begin{figure*}[t]
    \centering
    \includegraphics[width=0.9\linewidth, trim=18 18 18 18,clip]{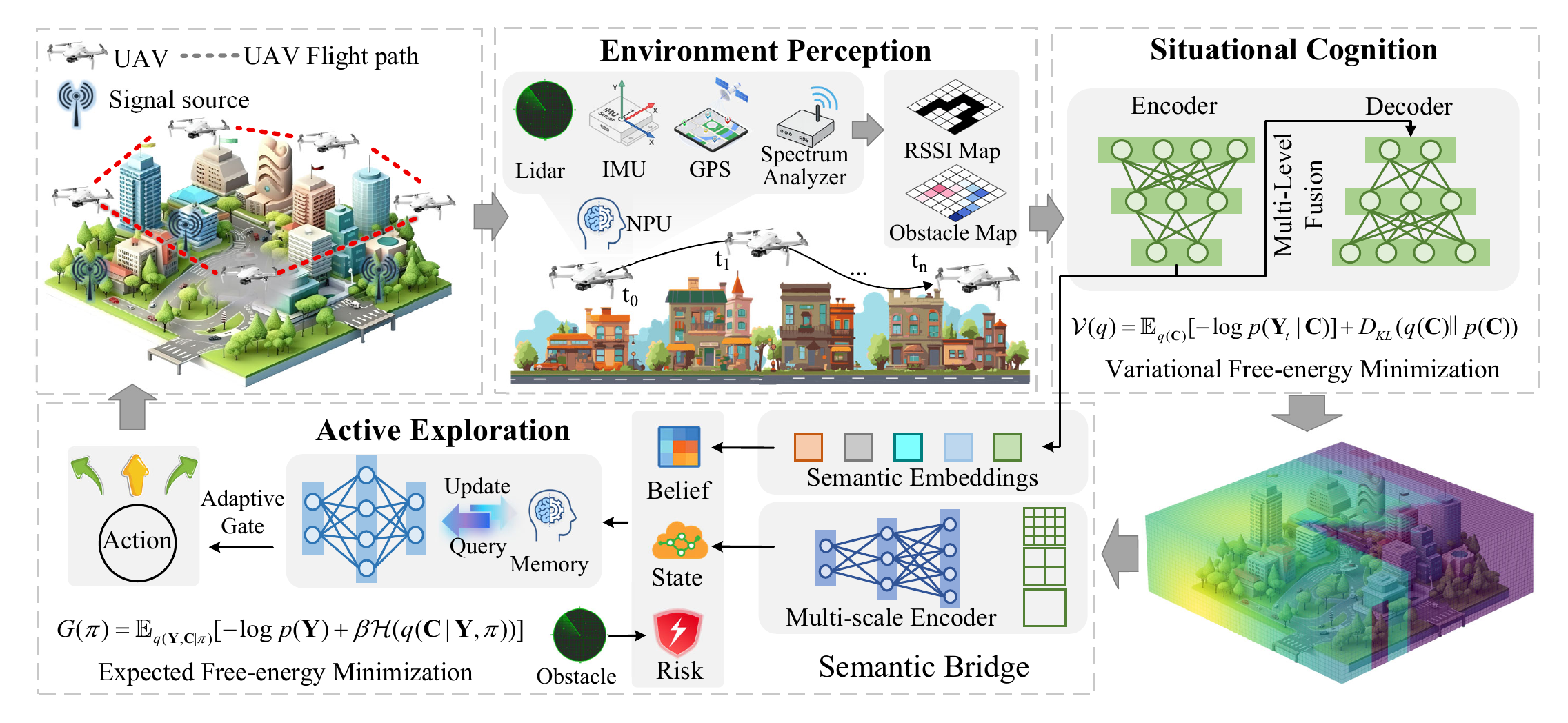}
    \caption{The proposed closed-loop framework integrating UAV perception, cognition, and active exploration for spectrum map construction.}

    \label{Fig.1}
\end{figure*}

\subsection{System Model}

Consider a UAV-based spectrum map construction system in a LAWN as illustrated in Fig.~\ref{Fig.1}. A UAV operates as an embodied agent within a three-dimensional monitoring region $\Omega \subset \mathbb{R}^3$ to construct an accurate spectrum map that captures the spatial distribution of RSSI across multiple frequency bands. To facilitate this, the continuous space is discretized into a regular grid structure $\mathcal{G} = \{(i, j, k) \mid 0 \leq i < N_x, 0 \leq j < N_y, 0 \leq k < N_z\}$ with resolutions $\Delta_x, \Delta_y, \Delta_z$, where $|\mathcal{G}| = N_x N_y N_z$. In this work the UAV surveys at a fixed altitude, so the evaluated grid reduces to the corresponding horizontal slice and the construction is performed over that slice for all monitored bands. Consequently, the ground truth spectrum map is represented as a complete multi-frequency field $\mathbf{C} \in \mathbb{R}^{|\mathcal{G}| \times M}$, where $M$ is the number of monitored frequency bands. The objective is to accurately construct the complete dense spectrum map $\hat{\mathbf{C}}$ from sparse measurements, minimizing the construction error between $\hat{\mathbf{C}}$ and $\mathbf{C}$ under an appropriate distance metric. Due to stringent SWaP constraints that severely limit flight endurance, the UAV can only collect extremely sparse measurements, with the total sampling budget bounded by $N_{budget} \ll |\mathcal{G}|$.

The spectrum field values are shaped by wireless propagation physics. Specifically, each entry $C_{i,f}$ of the map $\mathbf{C}$ represents the RSSI at grid cell $i \in \mathcal{G}$ and frequency $f \in \mathcal{F} = \{f_1, \dots, f_M\}$. This field is generated by one or more static transmitters, where the RSSI at position $\mathbf{p}$ and frequency $f$ follows the propagation model

\begin{equation}
P_{rx}(\mathbf{p}, f) = P_{tx}(f) - PL(d, f) - X_\sigma(\mathbf{p}) - L_{obs}(\mathbf{p}, f),
\label{eq:prx}
\end{equation}
where $P_{tx}(f)$ is the transmitter power, $PL(d, f)$ represents path loss over distance $d$ from the transmitter, $X_\sigma(\mathbf{p}) \sim \mathcal{N}(0, \sigma^2)$ models shadowing, and $L_{obs}(\mathbf{p}, f)$ accounts for obstacle attenuation.

\subsection{Framework Formulation}

As shown in Fig.~\ref{Fig.1}, the proposed framework operationalizes the EI paradigm through three tightly coupled components following a closed PCE loop. The loop begins at time step $t$ with the environment perception module, which aggregates multi-modal sensor measurements into a sparse measurement dataset $\mathbf{Y}_t$ and obstacle map $\mathbf{O}$. Next, the situational cognition module processes these inputs using a neural network $\mathcal{N}_{\psi}$ that embeds physics knowledge to construct a dense spectrum map estimate $\hat{\mathbf{C}}_t$, serving as the primary output representing the estimated spectrum distribution.
Simultaneously, this cognitive module extracts internal network representations from the encoder that capture spatial structure and sampling coverage, producing a semantic embedding $\mathbf{z}_{sem}^t \in \mathbb{R}^{d_{sem}}$ that encodes global construction status. These cognitive outputs are then fed to the active exploration module, which determines the next waypoint of the UAV based on the current position, the semantic embedding, and local map observations extracted from $\hat{\mathbf{C}}_t$. Executing this navigation command triggers a new sensing process that collects the RSSI measurement $\mathbf{m}_{t+1}$ at the new location and updates the dataset to $\mathbf{Y}_{t+1}$, thereby closing the loop. This bidirectional information flow between construction and navigation is formalized as

\begin{equation}
\mathbf{Y}_t \xrightarrow{\mathcal{N}_{\psi}} (\hat{\mathbf{C}}_t, \mathbf{z}_{sem}^t) \xrightarrow{\pi_\theta} a_t \xrightarrow{\mathcal{S}} \mathbf{Y}_{t+1},
\label{eq:info_flow}
\end{equation}
where $\mathcal{N}_{\psi}$ represents the cognition network that jointly outputs the spectrum map and semantic features, $\pi_\theta$ denotes the decision policy that selects navigation action $a_t$ from a discrete action space based on both global semantic embedding and local map observations, and $\mathcal{S}$ represents the sensing process that maps action $a_t$ to the new measurement $\mathbf{m}_{t+1}$ following the propagation model in eq.~\eqref{eq:prx}.

The nested observation history endows the closed loop with an information-state
property. Since $\mathbf{Y}_{t+1}=\mathbf{Y}_t\cup\{(\mathbf{p}_{t+1},\mathbf{m}_{t+1})\}$,
the $\sigma$-algebras satisfy $\sigma(\mathbf{Y}_t)\subseteq\sigma(\mathbf{Y}_{t+1})$,
so that, when these entropies are finite, the conditional entropy of the field is non-increasing on average,
\begin{equation}
h(\mathbf{C}\mid\mathbf{Y}_{t+1},\mathbf{O})\leq h(\mathbf{C}\mid\mathbf{Y}_t,\mathbf{O}),
\label{eq:entropy_monotonicity}
\end{equation}
where the gap equals the per-step information gain
$I(\mathbf{C};\mathbf{m}_{t+1}\mid\mathbf{Y}_t,\mathbf{p}_{t+1},\mathbf{O})\geq 0$. The policy therefore
controls the magnitude of this gain but not its sign. Furthermore, the decoupled
navigation state $(\mathbf{p}_t,N_t)$ is a measurable function of the augmented PCE
state $(\mathbf{p}_t,\theta_t,\mathbf{z}_{sem}^t,\hat{\mathbf{C}}_t^{local},N_t)$, so that every
policy defined on the former remains admissible on the latter and the optimal value
attainable on the augmented state is no smaller. The platform and simulator provide corresponding RSSI observations, obstacle geometry, and position variables. On the platform, these come from the spectrum analyzer, LiDAR, and GPS/IMU, respectively, and the simulator supplies them through its scene generator.

\subsubsection{Environment Perception}

As shown in Fig.~\ref{Fig.1}, the PCE loop begins with the environment perception, which fuses measurements from onboard sensors to produce two critical outputs for subsequent cognition. The first output is the sparse RSSI map represented as the accumulated measurement dataset $\mathbf{Y}_t$, and the second is the obstacle map $\mathbf{O}$ that defines navigable space and propagation barriers. Specifically, the LiDAR scans the environment to construct the obstacle map $\mathbf{O}: \mathcal{G} \to \{0, 1\}$, where $\mathbf{O}(\mathbf{p}) = 1$ indicates an obstacle at position $\mathbf{p}$ and $\mathbf{O}(\mathbf{p}) = 0$ denotes free space. This binary map serves dual purposes by identifying non-traversable regions for collision avoidance and providing structural priors for spectrum construction, as obstacles cause signal shadowing and diffraction that shape the spatial distribution of RSSI \cite{ZengWC2021_CKM,LaiSIGGRAPHAsia2024}. Moreover, navigation sensors including IMU and GPS provide accurate localization, yielding the UAV position $\mathbf{p}_t$ and heading $\theta_t$ at each timestep $t$.

The spectrum analyzer serves as the key sensing component that captures multi-frequency RSSI measurements. At position $\mathbf{p}_t$, the analyzer observes the RSSI vector $\mathbf{m}_t = [m_t^{(1)}, \dots, m_t^{(M)}]^T \in \mathbb{R}^M$ across the monitored frequency set $\mathcal{F}$. The relationship between the observed measurement $\mathbf{m}_t$ and the ground truth spectrum map $\mathbf{C}$ is formalized as
\begin{equation}
\mathbf{m}_t = \mathbf{C}_{\mathcal{I}(\mathbf{p}_t)} + \mathbf{n}_t,
\label{eq:measurement_model}
\end{equation}
where $\mathcal{I}(\mathbf{p}_t)$ maps the continuous position $\mathbf{p}_t$ to the discrete grid index $i \in \mathcal{G}$ through the nearest-neighbor assignment, $\mathbf{C}_i = [C_{i,f_1}, \dots, C_{i,f_M}]^T$ denotes the ground truth spectrum vector at grid cell $i$, and $\mathbf{n}_t = [n_t^{(1)}, \dots, n_t^{(M)}]^T \sim \mathcal{N}(\mathbf{0}, \sigma_n^2 \mathbf{I}_M)$ represents the measurement noise vector arising from hardware imperfections and environmental interference. The observation tuple $(\mathbf{p}_t, \mathbf{m}_t)$ is accumulated over time to form the sparse dataset $\mathbf{Y}_t = \mathbf{Y}_{t-1} \cup \{(\mathbf{p}_t, \mathbf{m}_t)\}$. Due to the stringent flight endurance limitation, the dataset remains extremely sparse with $N_t = |\mathbf{Y}_t| \ll |\mathcal{G}|$ throughout the mission. The sparse RSSI map provides limited measurements for the map construction, while the obstacle map serves dual purposes by enabling collision-free navigation and  providing structural priors for physics-informed construction through explicit obstacle geometry that supports learned representations of attenuation and shadowing.

\subsubsection{Situational Cognition}

\label{sec:situational_cognition}
The situational cognition module receives the sparse measurement dataset $\mathbf{Y}_t$ and obstacle map $\mathbf{O}$ from the perception module and aims to construct the complete dense spectrum map $\hat{\mathbf{C}} \in \mathbb{R}^{|\mathcal{G}| \times M}$ covering all spatial grid cells and frequency bands from the limited measurements. This construction task is inherently ill-posed because the number of measurements $N_t = |\mathbf{Y}_t|$ is significantly smaller than the number of unknown field values $|\mathcal{G}| \times M$, leading to infinitely many solutions that can fit the sparse data.

To address this ill-posedness, the construction is formulated as a Bayesian inference problem that seeks the posterior distribution $p(\mathbf{C} | \mathbf{Y}_t)$ representing the belief about the true field given the measurements~\cite{RomeroSPM2022,PolyzosTWC2024}. However, computing this posterior is computationally intractable due to the high-dimensional integral required to evaluate the evidence $p(\mathbf{Y}_t) = \int p(\mathbf{Y}_t|\mathbf{C})p(\mathbf{C})d\mathbf{C}$.  To overcome this intractability, we adopt the free energy principle (FEP) to formulate spectrum map construction from a Bayesian perspective using an approximate posterior $q(\mathbf C)$~\cite{IsomuraNC2023,FristonPhysRep2023}.  The variational free energy (VFE) is defined as

\begin{equation}
\mathcal{V}(q) = \mathbb{E}_{q(\mathbf{C})}[-\log p(\mathbf{Y}_t|\mathbf{C})] + D_{KL}(q(\mathbf{C})\|p(\mathbf{C})),
\label{eq:variational_free_energy}
\end{equation}
where $\mathbb{E}_{q(\mathbf{C})}$ denotes expectation over the approximate posterior, $p(\mathbf{Y}_t|\mathbf{C})$ is the likelihood function describing how measurements arise from the true field through eq.~\eqref{eq:measurement_model}, and $p(\mathbf{C})$ is the prior distribution encoding domain knowledge about plausible spectrum maps. Minimizing $\mathcal{V}(q)$ is equivalent to minimizing the KL divergence between the approximate posterior and true posterior $D_{KL}(q(\mathbf{C})\|p(\mathbf{C} | \mathbf{Y}_t))$~\cite{MillidgeNeuralComp2021WhenceEFE}.

The network $\mathcal{N}_{\psi}$ produces a single point estimate $\hat{\mathbf{C}}_t = \mathcal{N}_{\psi}(\mathbf{Y}_t, \mathbf{O})$.  The data-fidelity and prior terms motivate the following practical construction objective, formalized as 

\begin{equation}
\hat{\mathbf{C}}_t^* = \arg\min_{\hat{\mathbf{C}}} \mathcal{L}_{data}(\hat{\mathbf{C}}, \mathbf{Y}_t) + \lambda \mathcal{L}_{phys}(\hat{\mathbf{C}}),
\label{eq:construction_objective}
\end{equation}
where the data fidelity loss $\mathcal{L}_{data}(\hat{\mathbf{C}}, \mathbf{Y}_t) = \sum_{(\mathbf{p}_i, \mathbf{m}_i) \in \mathbf{Y}_t} \|\hat{\mathbf{C}}_{\mathcal{I}(\mathbf{p}_i)} - \mathbf{m}_i\|^2$ corresponds to the negative log-likelihood term and measures how well the constructed field explains the observed measurements, and the physics regularization $\mathcal{L}_{phys}(\hat{\mathbf{C}})$ approximates the negative log-prior by penalizing constructions that violate physics-informed priors. Moreover, the weighting parameter $\lambda$ balances data fidelity against physics-informed prior.

The motivation for introducing the physics-informed prior term is that purely data-driven methods relying solely on $\mathcal{L}_{data}$ may perform poorly under extreme sparsity. When $N_t \ll |\mathcal{G}| \times M$, the data fidelity term provides insufficient constraints and the network can produce infinitely many solutions that fit the sparse measurements but violate physics-informed priors, producing artifacts such as abrupt discontinuities or incorrect propagation patterns~\cite{WangTCCN2026RadioDiff}. To address this limitation, the prior $p(\mathbf{C})$ encodes wireless propagation physics from eq.~\eqref{eq:prx}, including distance-dependent power decay, frequency-selective fading, and obstacle shadowing effects.  The prior is realized through the encoder architecture, in which the ordered composition of frequency-related, distance-related, and spatial operations, together with the explicit obstacle and sampling-support inputs, biases the estimate toward physically plausible fields. Consequently, $\mathcal{L}_{phys}$ is not implemented as a stand-alone residual term, and its surrogate is provided jointly by this architectural bias and by the structural regularization used during training.

The primary output of this module is the constructed spectrum map $\hat{\mathbf{C}}_t \in \mathbb{R}^{|\mathcal{G}| \times M}$, which provides a complete spatial-frequency characterization of the spectrum map covering all grid cells and frequency bands. Beyond this primary construction output, the module also simultaneously extracts a semantic embedding $\mathbf{z}_{sem}^t \in \mathbb{R}^{d_{sem}}$ from the internal encoder representations that encodes global spatial patterns, measurement coverage distribution, and potentially informative regions.  This semantic feature summarizes the construction state through global spatial patterns and sampling coverage, providing contextual information for subsequent measurement selection.  Together, the spectrum map $\hat{\mathbf{C}}_t$ and semantic embedding $\mathbf{z}_{sem}^t$ are forwarded to the active exploration module to guide intelligent sampling decisions, thereby closing the PCE loop. The construction state therefore informs subsequent measurement selection, allowing the policy to adapt sampling decisions to the evolving map estimate and spatial coverage within the available budget.

\subsubsection{Active Exploration}

\label{sec:active_exploration}
The active exploration module closes the PCE loop by utilizing cognitive outputs from the situational cognition module to guide UAV navigation for the spectrum map construction. As shown in Fig.~\ref{Fig.1}, this module receives the semantic embedding $\mathbf{z}_{sem}^t$ and constructed spectrum map $\hat{\mathbf{C}}_t$ to determine the next action that collects informative measurements for improving construction quality. The core challenge is maximizing information gain for spectrum map construction under energy constraints. The UAV must collect measurements that reduce construction uncertainty, but flight time is limited by battery capacity. Inefficient trajectories consume the finite energy budget before sufficient measurements are gathered and therefore yield poor construction quality \cite{Zeng2016}. Therefore, the policy must balance informative sampling with navigation costs to support spectrum map construction within a limited action budget.

To formalize this balance, we adopt the expected free energy (EFE) from active inference~\cite{MillidgeNeuralComp2021WhenceEFE,IsomuraCommBio2022}.  An EFE-inspired reference objective for policy design can be written as 

\begin{equation}
G(\pi) = \mathbb{E}_{q(\mathbf{Y},\mathbf{C}|\pi)}[-\log p(\mathbf{Y}) + \beta \mathcal{H}(q(\mathbf{C}|\mathbf{Y},\pi))],
\label{eq:expected_free_energy}
\end{equation}
where $q(\mathbf{Y},\mathbf{C}|\pi)$ denotes the predicted distribution over future measurements and spectrum maps under policy $\pi$, $p(\mathbf{Y})$ encodes prior preferences over outcomes induced by the policy, which motivate navigation-related feedback in the practical reward, $\mathcal{H}(q(\mathbf{C}|\mathbf{Y},\pi))$ quantifies uncertainty in the constructed spectrum map, and $\beta$ balances these two terms. The first term reflects preferred outcomes associated with the navigation task. The second term favors observations that reduce expected uncertainty in the spectrum map. This objective guides the design of a policy that balances construction quality and navigation costs.

However, directly minimizing eq.~\eqref{eq:expected_free_energy} is computationally intractable because it requires integrating over possible future trajectories and measurements.  Therefore, active exploration is modeled as a Markov decision process (MDP) learned by DRL and guided by the one-step criterion, expressed as
\begin{equation}
\begin{aligned}
G_t(a_t)\approx{}&-\log p(a_t)\\
&+\beta\,\mathbb E_{q(\mathbf m_{t+1}\mid\mathbf Y_t,a_t)}
\left[\mathcal H\left(q(\mathbf C\mid\mathbf Y_{t+1},a_t)\right)\right].
\end{aligned}
\label{eq:instantaneous_free_energy}
\end{equation}
The first term represents action cost, while the second averages the uncertainty remaining after the next measurement. 

 The policy uses an augmented representation that combines navigation information with construction context. The tuple $\mathbf{s}_t^a=(\mathbf p_t,\theta_t,\mathbf z_{sem}^t,\hat{\mathbf C}_t^{local},N_t)$ includes position $\mathbf p_t$, heading $\theta_t$, semantic features $\mathbf z_{sem}^t$, local map observations $\hat{\mathbf C}_t^{local}$, and measurement count $N_t$. The policy uses this representation to select navigation actions.

 Construction error provides a practical training signal for the information-seeking objective in eq.~\eqref{eq:instantaneous_free_energy}. Local map observations and semantic features provide construction context for the policy. 
 During Stage-2 refinement, the reward reflects changes in construction accuracy and navigation progress. Let $E_t$ denote full-map RMSE and $D_t$ the navigation-distance signal from the training environment. The reward function can be formulated as
\begin{equation}
r_t=w_{\Delta}(E_{t-1}-E_t)-w_EE_t+w_D(D_{t-1}-D_t)+r_{aux},
\label{eq:reward_decomposition}
\end{equation}
where $w_{\Delta}$, $w_E$, and $w_D$ weight error reduction, current construction error, and navigation progress, respectively. The auxiliary term $r_{aux}$ includes a per-step time penalty, collision penalties, and feedback for successful or unsuccessful termination. This reward structure balances improvements in map construction with navigation progress and penalties for time, collisions, and terminal outcomes. 

We note that the ground-truth map $\mathbf{C}$ is required only during simulation-based policy training to evaluate eq.~\eqref{eq:reward_decomposition}, and in real-world deployment, the trained policy selects actions directly from the observed state $\mathbf{s}_t^{a}$ without any access to $\mathbf{C}$.

The policy $\pi_\theta$ is trained to maximize expected cumulative reward through environment interactions. Actions are sampled during training to enable exploration and selected greedily during deployment. Measurements are continuously collected along trajectories for spectrum map construction. Executing action $a_t$ triggers the perception module to update dataset $\mathbf{Y}_{t+1}$, which feeds the cognition module to produce updated map $\hat{\mathbf{C}}_{t+1}$ and features $\mathbf{z}_{sem}^{t+1}$, closing the PCE loop and enabling efficient construction under extreme sparsity and energy constraints.

Eq.~\eqref{eq:instantaneous_free_energy} expresses the balance between expected information gain and action cost over a single step within the EFE framework. In eq.~\eqref{eq:reward_decomposition}, map-error reduction represents observation value, while navigation, collision, time, and terminal terms capture movement and safety preferences. Together with the cognition objective in eq.~\eqref{eq:construction_objective}, these task-specific objectives connect data fidelity, prior knowledge, and construction-aware exploration within the FEP framework. The VFE and EFE expressions guide their design and interpretation without requiring explicit evaluation of posterior entropy during training.

\begin{figure*}[t]
    \centering
    \includegraphics[width=0.9\linewidth, trim=18 18 18 18,clip]{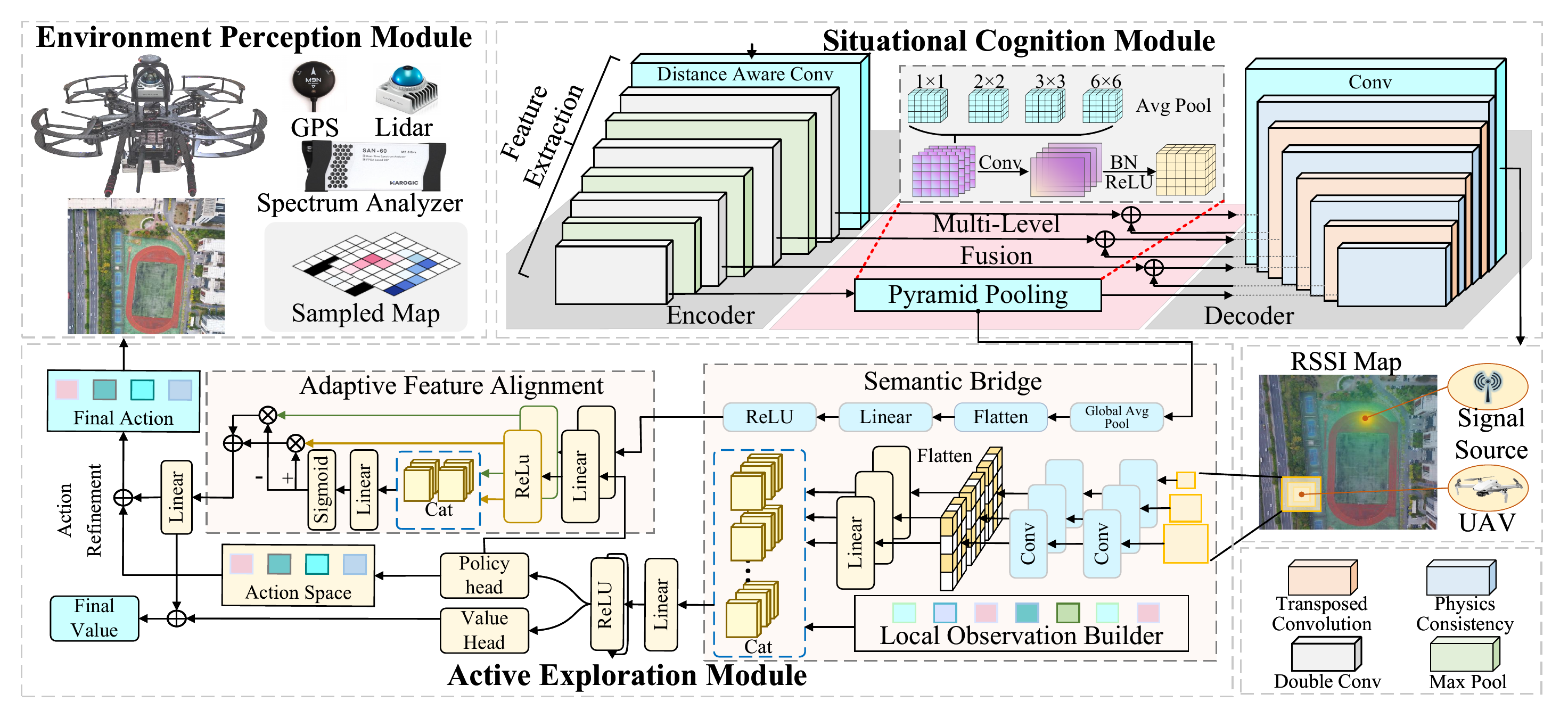}
    \caption{Architecture of the ECDNet comprising environment perception, physics-informed situational cognition, and active exploration modules.}

    \label{Fig.2}
\end{figure*}

\subsection{Optimization Objective}

The framework optimization addresses a spectrum map construction problem that co-optimizes the UAV trajectory $\boldsymbol{\tau} = \{\mathbf{p}_0, \mathbf{p}_1, \ldots, \mathbf{p}_T\}$ and the spectrum map estimate $\hat{\mathbf{C}}$ to minimize construction error while respecting the sparse sampling constraint. This joint optimization is formalized as

\begin{subequations}
\label{eq:joint_objective_new}
\begin{align}
\min_{\boldsymbol{\tau}, \psi, \theta} \quad & \mathbb{E}_{\mathbf{C} \sim p_{env}}[\|\mathcal{N}_\psi(\mathbf{Y}(\boldsymbol{\tau}), \mathbf{O}) - \mathbf{C}\|^2] \label{eq:joint_main} \\
\text{s.t.} \quad & \begin{gathered}[t]
\boldsymbol{\tau} = \{\mathbf{p}_t\}_{t=0}^T,\\
{\mathbf p_{t+1}=\mathbf p_t+\Delta\mathbf p(a_t,\theta_t),\quad 0\leq t<T}
\end{gathered}, \label{eq:trajectory_constraint} \\
& a_t = \pi_\theta(\mathbf{s}_t^a), \quad \mathbf{s}_t^a = (\mathbf{p}_t,\theta_t,\mathbf{z}_{sem}^t,\hat{\mathbf{C}}_t^{local},N_t), \label{eq:policy_constraint} \\
& N_T \le N_{budget}, \quad \mathbf{p}_t \notin \mathbf{O} \quad \forall t, \label{eq:joint_practical}
\end{align}
\end{subequations}
where the main objective eq.~\eqref{eq:joint_main} minimizes the construction error over environment instances $\mathbf{C}$ sampled from the environment distribution $p_{env}$, with the measurement dataset $\mathbf{Y}(\boldsymbol{\tau}) = \{(\mathbf{p}_t, \mathbf{m}_t)\}_{t=0}^T$ explicitly dependent on the trajectory $\boldsymbol{\tau}$. This formulation captures the bidirectional coupling where the trajectory determines measurement locations that define the dataset, which in turn determines the map quality and extracted cognitive features that guide future trajectory decisions through the augmented state $\mathbf{s}_t^a$.

The constraints in eq.~\eqref{eq:trajectory_constraint} and eq.~\eqref{eq:policy_constraint} formalize how the trajectory is generated through sequential actions selected by the policy $\pi_\theta$ operating on the augmented state containing cognitive features from the construction network.  The policy maximizes expected cumulative reward using the construction and navigation terms in eq.~\eqref{eq:reward_decomposition}.  The practical constraints in eq.~\eqref{eq:joint_practical} enforce the sampling budget and collision avoidance requirements.  This formulation couples trajectory planning with map construction at the task level. The networks are trained sequentially as described in Algorithm~1.

\section{Our Proposed ECDNet}

This section presents the detailed architecture design of the ECDNet illustrated in Fig.~\ref{Fig.2}, which instantiates the proposed framework through three tightly coupled modules: environment perception, physics-informed situational cognition, and active exploration.

\subsection{Environment Perception Module}

The environment perception module serves as the ECDNet interface to the physical world, converting the sparse measurement dataset $\mathbf{Y}_t$ and obstacle map $\mathbf{O}$ into a fixed-size tensor representation for neural network processing. As shown in Fig.~\ref{Fig.2}, the module constructs a multi-channel input tensor $\mathbf{X}^{in}$ that encapsulates all information for the cognition network.

The input tensor $\mathbf{X}^{in}$ is constructed from four complementary components.
The sparse sampled map $\mathbf{I}_{sparse}$ carries the measured RSSI of the monitored
bands at their grid locations and leaves unmeasured cells zero-filled. The interpolated
RSSI channels $\mathbf{I}_{interp}$ provide dense provisional fields for the same bands.
A binary sampling mask $\mathbf{M}_{sam}$ distinguishes observed cells from zero-filled
unobserved cells and records measurement support. The obstacle map $\mathbf{O}$ provides
geometry for propagation-aware construction and motion feasibility. The complete input
is
\begin{equation}
\mathbf{X}^{in} = \operatorname{Concat}(\mathbf{I}_{sparse},\mathbf{I}_{interp},\mathbf{M}_{sam},\mathbf{O}),
\label{eq:input_tensor}
\end{equation}
in which the sparse and the interpolated channels provide measured evidence and
provisional spatial context, the sampling mask preserves observation provenance, and the
obstacle map supplies explicit environmental geometry.

\subsection{Situational Cognition Module}

\begin{figure}[t]
\hspace*{-0cm}
\centering
\includegraphics[width=6.5cm,trim=0 0 0 0,clip]{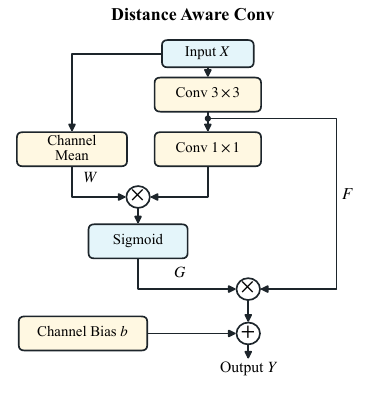}
\caption{Internal structure of the distance-aware convolution layer with input-conditioned spatial weighting and a learned channel bias.}

\label{Fig.3}
\end{figure}

The situational cognition module implements the cognition network $\mathcal{N}_\psi$, which transforms $\mathbf{X}^{in}$ into a dense spectrum-map estimate $\hat{\mathbf{C}}_t$. Its physics-informed prior is distributed across the complete encoder. The encoder implements a physics-informed transformation

\begin{equation}
{\Phi_{phys}=\Phi_{spatial}\circ\Phi_{dist}\circ\Phi_{freq},}
\label{eq:physics_composition}
\end{equation}
 where $\Phi_{dist}$ applies input-conditioned spatial weighting, $\Phi_{freq}$ models frequency-specific responses, and $\Phi_{spatial}$ aggregates spatial context.   As shown in Fig.~\ref{Fig.3}, distance-aware convolutions use input-conditioned spatial weights and learned channel biases alongside frequency-specific sub-networks and attention mechanisms. The distance-aware layer receives no explicit transmitter-distance coordinates and imposes no analytical path-loss law.   Obstacle geometry and measurement support enter explicitly through $\mathbf{O}$ and $\mathbf{M}_{sam}$, while obstacle-induced shadowing is learned from the controlled training distribution motivated by eq.~\eqref{eq:prx}. The resulting architecture provides an inductive bias toward physically plausible spectrum maps.

The complete forward pass processes $\mathbf{X}^{in}$ through three stages as shown in Fig.~\ref{Fig.2}. The encoder generates feature maps and skip-connection tensors. A pyramid pooling module then aggregates multi-scale context from the encoder features. Finally, the decoder fuses the pooled context with skip connections to construct the map. This process is formalized as

\begin{subequations}
\label{eq:network_forward}
\begin{align}
\mathbf{Z}_{enc}, \mathbf{Z}_{skip} &= \mathcal{E}_\psi(\mathbf{X}^{in}), \label{eq:encoder} \\
\mathbf{Z}_{pool} &= \mathcal{P}_{pool}(\mathbf{Z}_{enc}), \label{eq:pooling} \\
\hat{\mathbf{C}}_t &= \mathcal{D}_\psi(\mathbf{Z}_{pool}, \mathbf{Z}_{skip}), \label{eq:decoder}
\end{align}
\end{subequations}
where $\mathcal{E}_\psi$, $\mathcal{P}_{pool}$, and $\mathcal{D}_\psi$ denote the encoder, pyramid pooling, and decoder components, respectively, $\mathbf{Z}_{enc}$ represents encoder output features, $\mathbf{Z}_{pool}$ represents pooled context, and $\mathbf{Z}_{skip}$ denotes multi-level skip-connection tensors.

 The network is trained with a composite loss motivated by the construction objective in eq.~\eqref{eq:construction_objective}. This loss, which combines three complementary terms for data fitting and structural regularization, is formulated as 

\begin{subequations}
\label{eq:cognitive_loss}
\begin{align}
\mathcal{L}_{cog}(\psi) = & \alpha \mathbb{E}_{\mathbf{p} \in \mathcal{G}}[\mathbf{M}_{sam}(\mathbf{p}) \cdot \|\hat{\mathbf{C}}(\mathbf{p}) - \mathbf{C}(\mathbf{p})\|^2] \label{eq:loss_data} \\
+ & \mathbb{E}_{\mathbf{p} \in \mathcal{G}}[\|\hat{\mathbf{C}}(\mathbf{p}) - \mathbf{C}(\mathbf{p})\|^2] \label{eq:loss_recon} \\
+ & \lambda_{perc} \sum_{l} \|\Phi_l^{VGG}(\hat{\mathbf{C}}) - \Phi_l^{VGG}(\mathbf{C})\|_1, \label{eq:loss_phys}
\end{align}
\end{subequations}
where $\psi$ denotes trainable parameters, $\alpha$ is a large weight emphasizing sampled locations, $\mathbf{M}_{sam}$ is the binary sampling mask from eq.~\eqref{eq:input_tensor}, $\lambda_{perc}$ controls the structural regularization strength, and $\Phi_l^{VGG}$ extracts features from the $l$-th layer of a pretrained VGG network.  The first term in eq.~\eqref{eq:loss_data} provides sampled-location supervision by assigning weight $\alpha$ to cells indicated by $\mathbf{M}_{sam}$. It approximates the data-fidelity role of eq.~\eqref{eq:construction_objective} during offline training without imposing exact agreement with noisy field measurements. 

The second term eq.~\eqref{eq:loss_recon} provides supervised global construction guidance during offline training, while the third term eq.~\eqref{eq:loss_phys} acts as a multi-scale structural regularizer that suppresses spatially incoherent artifacts.  The physics prior in $\mathcal{L}_{phys}$ is represented by propagation-motivated feature processing across the network, while the structural regularizer constrains the constructed output.  These architectural and loss-based components provide the surrogate for the conceptual prior in eq.~\eqref{eq:construction_objective}.

Beyond producing $\hat{\mathbf{C}}_t$, the module extracts the semantic embedding
$\mathbf{z}_{sem}^t=\Psi_{extract}(\mathbf{Z}_{enc})$ from the global bottleneck
representation through pyramid pooling and global average pooling. Because $\mathbf{M}_{sam}$ is an explicit input, observation support and
coverage are encoded in $\mathbf{Z}_{enc}$ and can be transferred to the policy. The
local crop $\hat{\mathbf{C}}_t^{local}$ supplies immediate neighborhood detail, whereas
$\mathbf{z}_{sem}^t$ supplies a complementary global summary of the construction state,
including coverage, large-scale spectrum variation, and obstacle-related structure.

\begin{algorithm}[t]
\caption{Sequential Training Procedure}

\label{alg:training}
\begin{algorithmic}[1]
\State \textbf{Input:} Offline dataset $\mathcal{D}_{train}$, parallel environments $\{Env_k\}_{k=1}^K$

\State \textbf{Output:} Frozen networks $\mathcal{N}_{\psi^*}$, $\pi^{(1)}_{\theta_1^*}$, refined policy $\pi_{refine}$

\State \textit{// Cognition Network Training}

\For{$epoch = 1, \ldots, N_{cog}$}

    \State Sample minibatch $\{(\mathbf{Y}, \mathbf{O}, \mathbf{C})\}$ from $\mathcal{D}_{train}$

    \State $\hat{\mathbf{C}} \leftarrow \mathcal{N}_\psi(\mathbf{Y}, \mathbf{O})$ using eq.~\eqref{eq:network_forward}

    \State $\psi \leftarrow \psi - \nabla_\psi \mathcal{L}_{cog}(\psi)$ using eq.~\eqref{eq:cognitive_loss}

\EndFor

\State Save $\psi^*$ and freeze

\State \textit{// Baseline Policy Training}

\State Initialize $\pi^{(1)}_{\theta_1}$ and $V_{\phi_1}$

\For{$iter = 1, \ldots, N_{nav}$}

    \For{each environment $k \in \{1, \ldots, K\}$}

        \State Collect trajectory using $\pi^{(1)}_{\theta_1}$ and $\mathcal{N}_{\psi^*}$

        \State Compute $r_t^{(1)}$ from map-error improvement, distance, safety, and auxiliary terms

    \EndFor

    \State Compute advantages $\{\hat{A}_t\}$ from all trajectories

    \State $(\theta_1, \phi_1) \leftarrow$ update using PPO

\EndFor

\State Save $\theta_1^*$ and freeze

\State \textit{// Policy Network Refinement}

\State Initialize refiner heads on frozen $\pi^{(1)}_{\theta_1^*}$ and $\mathcal{N}_{\psi^*}$

\For{$iter = 1, \ldots, N_{refine}$}

    \For{each environment $k \in \{1, \ldots, K\}$}

        \State Collect trajectory using $\pi_{refine}$

        \For{each state in trajectory}

            \State Extract $\hat{\mathbf{C}}_t$, $\mathbf{z}_{sem}^t$, and $\hat{\mathbf{C}}_t^{local}$ from frozen $\mathcal{N}_{\psi^*}$

            \State $\mathbf{s}_t^{a} \gets (\mathbf{p}_t, \theta_t, \mathbf{z}_{sem}^t, \hat{\mathbf{C}}_t^{local}, N_t)$

        \EndFor

        \State Compute $r_t$ from eq.~\eqref{eq:reward_decomposition}

    \EndFor

    \State Compute advantages $\{\hat{A}_t\}$ from all trajectories

    \State Update $\theta_{refine}$ via PPO

\EndFor

\State \textbf{Return} $\mathcal{N}_{\psi^*}$, $\pi^{(1)}_{\theta_1^*}$, $\pi_{refine}$

\end{algorithmic}
\end{algorithm}

\subsection{Active Exploration Module}

\label{sec:active_exploration_module}

The active exploration module implements the policy $\pi_\theta$ that maps the augmented state $\mathbf{s}_t^{a}$ to navigation actions $a_t$ for spectrum map construction. In implementation, $\pi_\theta$ is realized in two stages.  As shown in Fig.~\ref{Fig.2}, the Stage-1 baseline $\pi^{(1)}_{\theta_1}$ uses a feedforward policy for navigation in the evaluated configuration. The Stage-2 refiner freezes both $\pi^{(1)}_{\theta_1^*}$ and $\mathcal{N}_{\psi^*}$ and learns residual corrections to the Stage-1 logits and value estimate, with small initial residual-head weights.

The module architecture comprises two complementary policy stages with distinct input scopes. The Stage-1 baseline policy focuses on stable navigation from physical observations and local constructed-map observations. Specifically, it uses a local processing pathway $\Phi_{local}$ that encodes the physical position $\mathbf{p}_t$, heading $\theta_t$, local map observations $\hat{\mathbf{C}}_t^{local}$ extracted around the current position from the constructed map $\hat{\mathbf{C}}_t$, obstacle map $\mathbf{O}$, and measurement count $N_t$ into a local feature vector $\mathbf{z}_{local}$. This local pathway processes the constructed RSSI field and its spatial consistency patterns to represent local construction state. The Stage-1 input is therefore

\begin{equation}
\mathbf{z}^{(1)}_{t} = \Phi_{local}(\mathbf{p}_t, \theta_t, \hat{\mathbf{C}}_t^{local}, \mathbf{O}, N_t),
\label{eq:semantic_bridge_fusion}
\end{equation}
which gives the baseline navigation policy local observations without directly using the global semantic embedding. The Stage-2 refiner then introduces the semantic bridge. A global processing pathway $\Phi_{global}$ encodes the cognition-derived semantic embedding $\mathbf{z}_{sem}^t$ into a global feature vector that summarizes construction-related spatial patterns and global observation context. This global feature is combined with the frozen Stage-1 logits through the gated residual mechanism described below, so that the global semantic embedding enters the policy only at the refinement stage.

The second component is the policy network. The Stage-1 baseline policy $\pi^{(1)}_{\theta_1}$ uses feedforward layers to process $\mathbf{z}^{(1)}_{t}$ and outputs the baseline action logits $\mathbf{l}^{(1)}_t$ and value $v^{(1)}_t$.
The Stage-2 refiner freezes both $\pi^{(1)}_{\theta_1^*}$ and $\mathcal{N}_{\psi^*}$, and refines the baseline logits by adaptively mixing the frozen policy representation with the cognition-derived representation:

\begin{subequations}
\label{eq:adaptive_alignment}
\begingroup
\setlength{\jot}{1pt}
\begin{align}
(\mathbf{l}^{(1)}_t,v^{(1)}_t)
&= \pi^{(1)}_{\theta_1^*}(\mathbf{z}^{(1)}_{t}), \\
\mathbf{z}_{sem}^t &= \Psi_{extract}(\mathbf{Z}_{enc}), \\
\mathbf{r}^{stage}_t &= \Phi_{stage}\big(\mathrm{sg}(\mathbf{l}^{(1)}_t)\big), \\
\mathbf{r}^{enc}_t &= \Phi_{enc}(\mathbf{z}_{sem}^t), \\
\mathbf{g}_t &= \sigma\big(\mathrm{MLP}_g([\mathbf{r}^{stage}_t,\mathbf{r}^{enc}_t])\big), \\
\mathbf{u}_t &= \mathbf{g}_t \odot \mathbf{r}^{stage}_t
+ (1-\mathbf{g}_t)\odot \mathbf{r}^{enc}_t, \\
\mathbf{l}_{final} &= \mathrm{sg}(\mathbf{l}^{(1)}_t)
+ \Delta_{\pi}(\mathbf{u}_t), \label{eq:final_action}\\
v_{final} &= \mathrm{sg}(v^{(1)}_t) + V_{\Delta}(\mathbf{u}_t).
\end{align}
\endgroup
\end{subequations}
where $\mathbf{Z}_{enc}$ is extracted from the frozen cognition encoder, $\Phi_{stage}$ and $\Phi_{enc}$ denote the stage branch and encoder branch, respectively, and $\mathbf{g}_t$ is an element-wise gate conditioned on both representations. Thus, $\mathbf{u}_t$ adaptively mixes the frozen policy embedding and the cognition-derived embedding before producing the residual policy correction. The stop-gradient operator $\mathrm{sg}(\cdot)$ indicates that the Stage-1 policy is not updated during refinement. Similar to the policy logits, the value estimate is formulated as a residual correction on top of the frozen Stage-1 value head.  No recurrent hidden state is used in the evaluated configuration. Both stages are trained with PPO, with Stage~1 updating the feedforward policy and value head and Stage~2 updating only the refiner heads.  The PPO objective to be maximized can be written as

\begin{equation}
\mathcal{L}_{dec}(\theta, \phi) = \mathbb{E}_t[\mathcal{L}_{CLIP}(\theta) - c_1 \mathcal{L}_{VF}(\phi) + c_2 \mathcal{S}_\pi(\theta)],
\label{eq:ppo_objective}
\end{equation}
where $\mathcal{L}_{CLIP}$, $\mathcal{L}_{VF}$, and $\mathcal{S}_\pi$ denote the clipped surrogate objective, value loss, and entropy bonus, respectively.

In practice, the local map observations $\hat{\mathbf{C}}_t^{local}$ are extracted by cropping a local region centered at the current position $\mathbf{p}_t$ from the full constructed map $\hat{\mathbf{C}}_t$ and processing it through multi-scale convolutional layers. This local observation encodes the constructed RSSI and its spatial patterns, providing learned neighborhood context for action selection during spectrum map construction. This module completes the closed-loop framework. The situational cognition module constructs spectrum maps and extracts semantic embeddings. The active exploration module uses these embeddings along with local map observations to select actions conditioned on the current construction state and spatial coverage. Executing action $a_t$ triggers the environment perception module to collect measurement $\mathbf{m}_{t+1}$ and update the dataset, thereby restarting the PCE loop and updating the constructed map as additional measurements are collected through embodied exploration.

\begin{algorithm}[t]
\caption{Real-Time Deployment Procedure}

\label{alg:deployment}
\begin{algorithmic}[1]
\Require Networks $(\mathcal{N}_{\psi^*}, \pi_{policy})$, obstacle map $\mathbf{O}$, step budget $T_{budget}$, configured minimum coverage $\rho_{min}$

\Ensure Constructed map $\hat{\mathbf{C}}_{final}$, trajectory $\boldsymbol{\tau}$

\State Deploy UAV at initial position $\mathbf{p}_0$ with heading $\theta_0$

\State $\mathbf{Y}_{-1} \gets \emptyset$, $N_{-1} \gets 0$, $t \gets 0$, $\boldsymbol{\tau} \gets (\mathbf{p}_0)$

\State $\mathcal{G}_{free} \gets \{\mathbf{p} \in \mathcal{G} \mid \mathbf{O}(\mathbf{p}) = 0\}$

\While{$t < T_{budget}$}

    \State $\mathbf{m}_t \gets$ measure RSSI at $\mathbf{p}_t$

    \State $\mathbf{Y}_t \gets \mathbf{Y}_{t-1} \cup \{(\mathbf{p}_t, \mathbf{m}_t)\}$, $N_t \gets N_{t-1} + 1$

    \State $\hat{\mathbf{C}}_t \gets \mathcal{N}_{\psi^*}(\mathbf{Y}_t, \mathbf{O})$ using eq.~\eqref{eq:network_forward}

    \State $\hat{\mathbf{C}}_{final} \gets \hat{\mathbf{C}}_t$

    \State $\hat{\mathbf{C}}_t^{local} \gets$ local region around $\mathbf{p}_t$ from $\hat{\mathbf{C}}_t$

    \If{$\pi_{policy} = \pi_{refine}$}
        \State $\mathbf{Z}_{enc}, \mathbf{Z}_{skip} \gets$ outputs from eq.~\eqref{eq:encoder}

        \State $\mathbf{z}_{sem}^t \gets \Psi_{extract}(\mathbf{Z}_{enc})$

    \Else
        \State $\mathbf{z}_{sem}^t \gets \mathbf{0}_{d_{sem}}$

    \EndIf
    \State $\mathbf{s}_t^{a} \gets (\mathbf{p}_t, \theta_t, \mathbf{z}_{sem}^t, \hat{\mathbf{C}}_t^{local}, N_t)$

    \If{$\pi_{policy} = \pi_{refine}$}
        \State $a_t \gets \pi_{refine}(\mathbf{s}_t^{a})$

    \Else
        \State $a_t \gets \pi^{(1)}_{\theta_1^*}(\mathbf{s}_t^{a})$

    \EndIf
    \State Update $(\mathbf{p}_{t+1},\theta_{t+1})$ from $(\mathbf{p}_t,\theta_t)$ using $a_t$

    \State Append $\mathbf{p}_{t+1}$ to $\boldsymbol{\tau}$

    \If{$|\{\mathbf{p}\mid(\mathbf{p},\mathbf{m})\in\mathbf{Y}_t\}| / |\mathcal{G}_{free}| \ge \rho_{min}$}
        \State \textbf{break}

    \EndIf
    \State $t \gets t + 1$

\EndWhile

\State \textbf{Return} $\hat{\mathbf{C}}_{final}$, $\boldsymbol{\tau}$

\end{algorithmic}
\end{algorithm}

\subsection{Training and Deployment Strategy}

The framework employs a three-stage training strategy, as summarized in Algorithm~\ref{alg:training}. First, the cognition network $\mathcal{N}_{\psi}$ is pretrained on $\mathcal{D}_{train}$ by minimizing eq.~\eqref{eq:cognitive_loss}, and the converged parameters $\psi^*$ are frozen. Second, the Stage-1 baseline policy $\pi^{(1)}_{\theta_1}$ is trained with PPO using a baseline reward that combines preliminary map-error improvement, distance efficiency, safety cost, and auxiliary terms. The resulting $\theta_1^*$ is frozen as a stable navigation backbone. Third, the Stage-2 refiner is trained on top of the frozen cognition network and frozen baseline policy, where only the refiner heads are updated using the full reward in eq.~\eqref{eq:reward_decomposition}. This curriculum keeps the baseline parameters fixed while learning construction-aware residual corrections to the policy logits.

During deployment detailed in Algorithm~\ref{alg:deployment}, the framework operates without ground truth, with $\mathcal{G}_{free}$ representing navigable cells. At each timestep, the UAV measures spectrum $\mathbf{m}_t$, constructs $\hat{\mathbf{C}}_t$ using frozen $\mathcal{N}_{\psi^*}$, and forms $\mathbf{s}_t^{a}$ from local observations and semantic features. The selected policy then computes action $a_t$.  A rollout terminates either at the step budget $T_{budget}$ or at the configured coverage threshold $\rho_{min}$. Coverage counts distinct sampled free cells, with revisits contributing no additional coverage. Because revisits are permitted, the step budget upper-bounds the number of distinct sampled cells collected in one rollout.

\section{Experimental Results}

\begin{table*}[t]
\centering

\caption{Simulation and deployment platform specifications. }

\label{tab:platform_specs}
\setlength{\tabcolsep}{12pt}
\renewcommand{\arraystretch}{1.15}
\begin{tabular}{@{}llll@{}}
\toprule
\multicolumn{2}{c}{\textbf{Simulation Platform}} & \multicolumn{2}{c}{\textbf{Deployment Platform}} \\
\cmidrule(r){1-2}\cmidrule(l){3-4}
Operating system & Ubuntu 22.04 LTS & Hardware & Jetson Orin NX, 16 GB \\
GPU & NVIDIA RTX 4090 & GPU cores & 1024 CUDA \\
CPU & Intel i9-13900K & Cognition latency & 27.63 ms (p95 28.22) \\
RAM & 64 GB DDR5 & Decision latency & 8.16 ms (p95 9.07) \\
Framework & PyTorch 2.0.1 & Masking and action generation & 1.54 ms (p95 1.70) \\
Python & 3.10.12 & Model-loop latency & 37.33 ms (p95 38.48) \\
CUDA & 12.1 & Timing scope & GPU-resident model loop \\
Batch size & 32 & Precision and batch & FP32, batch 1 \\
Learning rates & 1e-4 and 3e-4 & Power mode & 25 W \\
Optimizer & Adam & Control period & 100 ms (10 Hz) \\
& & Endurance & 18 min \\
\bottomrule
\end{tabular}
\end{table*}
\begin{figure}[t]
\centering
\includegraphics[width=8cm,trim=5 5 5 5,clip]{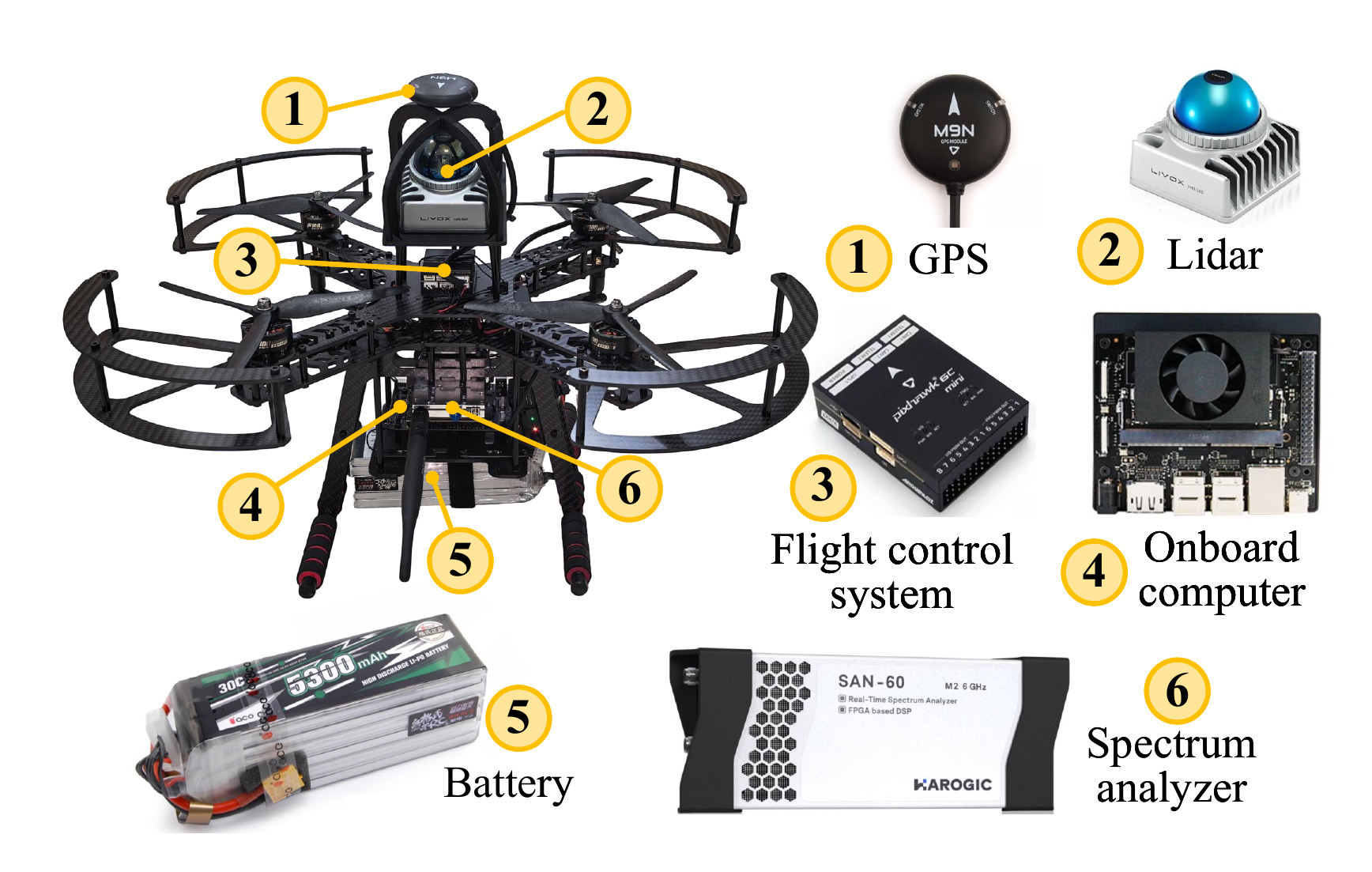}
\caption{Hardware architecture of the developed UAV-based embodied spectrum construction platform with integrated sensing and computing modules.}

\label{Fig.4}
\end{figure}

\begin{figure}[t]
\centering
\includegraphics[width=8cm,trim=1 1 1 1,clip]{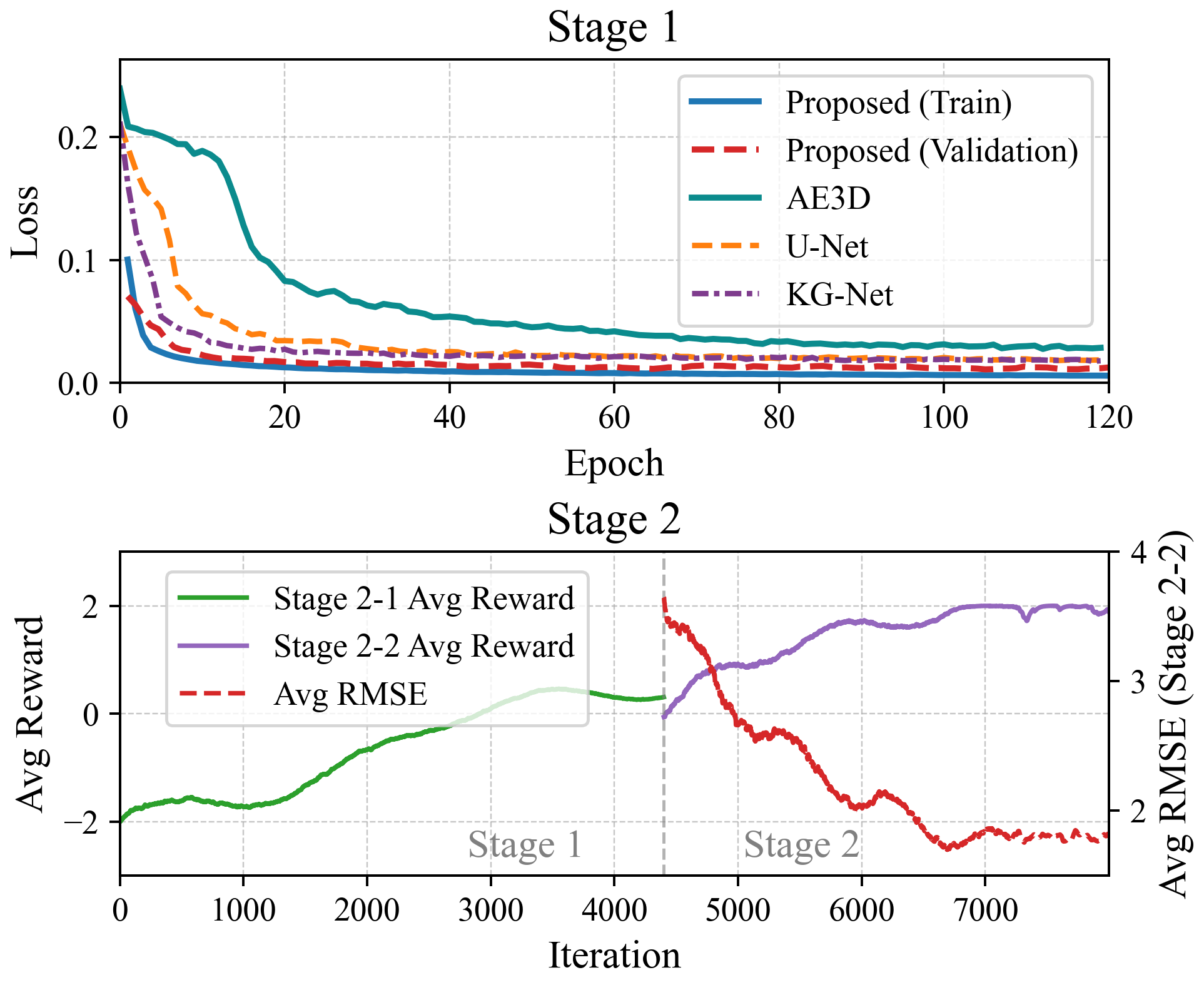}
\caption{Convergence analysis of cognition-network pretraining and policy-network
learning under the sequential training strategy. In the lower panel, the two reward
traces labeled Stage 2-1 and Stage 2-2 correspond to Stage-1 and Stage-2 policy
training, and the RMSE trace corresponds to the Stage-2 evaluation of the refined policy.}

\label{Fig.5}
\end{figure}

\begin{figure*}[!t]
\centering
\includegraphics[width=0.86\linewidth,trim=5 5 5 5,clip]{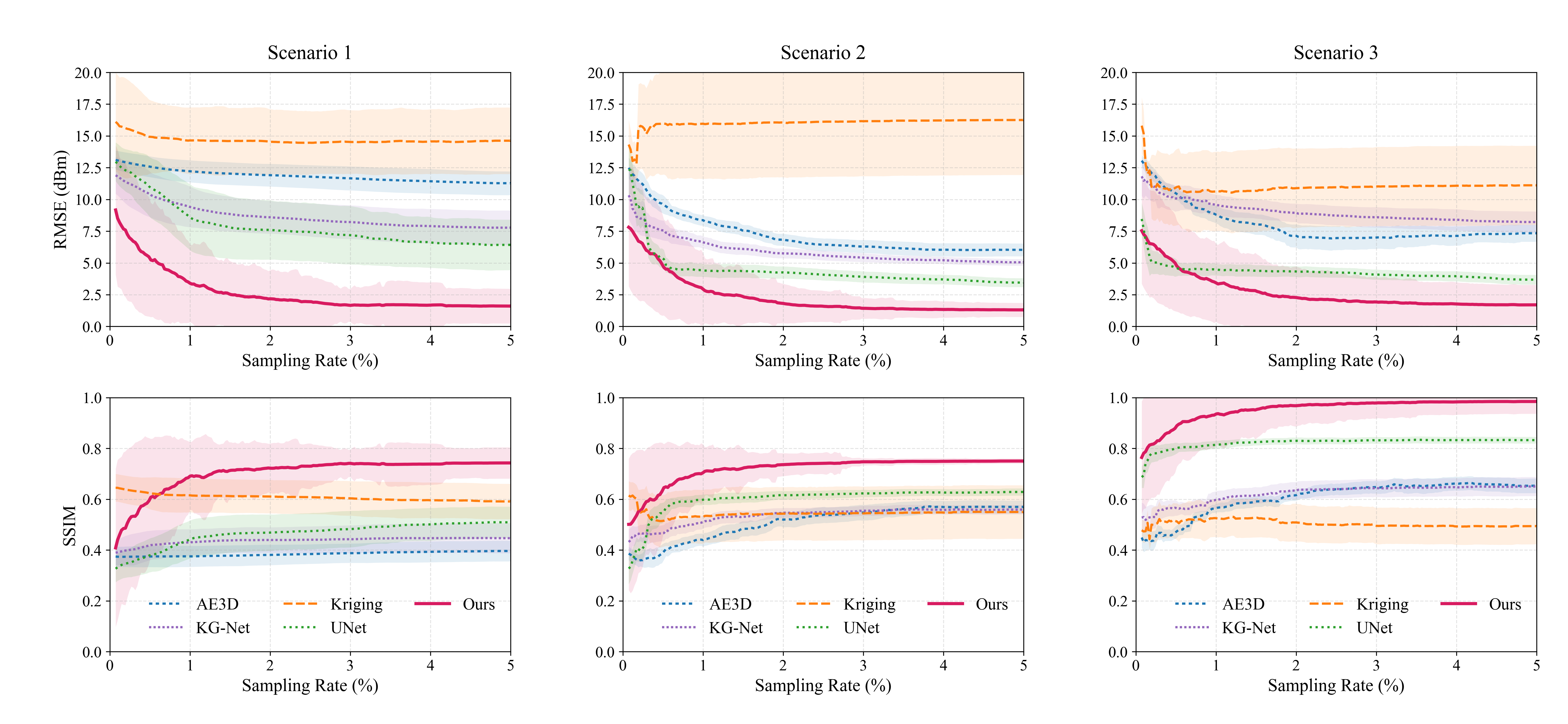}
\caption{Quantitative comparison of RMSE and SSIM under sparse sampling. Scenario~1, Scenario~2, and Scenario~3 contain 186, 189, and 181 held-out spectrum-map instances, respectively. At each sampling level, curves report the mean across maps in each scenario, and shaded regions denote $\pm1$ sample standard deviation across maps rather than across independently trained runs. The plotted axis is a continuous sampling-rate sweep and is not restricted to the six operating points of Table~II.}

\label{Fig.6}
\end{figure*}

This section presents a comprehensive evaluation of the proposed ECDNet.
The assessment spans controlled simulations and a prototype-level
demonstration on an embedded UAV platform. To validate the feasibility of
autonomous spectrum map construction under strict hardware constraints, we developed the
custom UAV platform illustrated in Fig.~\ref{Fig.4}. As shown in Fig.~\ref{Fig.4},
the system integrates a Livox Mid-360 LiDAR and a HAROGIC SAN-60 M2 spectrum analyzer
within a compact airframe. A Jetson Orin NX processes the data onboard. However,
deploying complex EI algorithms on such devices constrained by size, weight, and
power (SWaP) poses significant computational challenges due to hardware
limitations.

Table~\ref{tab:platform_specs} reports the detailed system efficiency metrics.
 The Jetson latencies are measured with GPU-resident inputs, FP32 inference,
and CUDA synchronization at every timing boundary. The cognition, decision, and
action-masking stages require 27.63, 8.16, and 1.54~ms on average, and the complete
model loop requires only 37.33~ms (p95 38.48~ms), which is less than 40\% of the
configured 100~ms control period. These results demonstrate that the complete
cognition--decision chain of ECDNet can be executed onboard within a single control
period on the SWaP-constrained edge device.

Beyond the hardware demonstration above, we conduct comprehensive simulation
experiments in three synthetic urban environments with increasing propagation
complexity to systematically evaluate construction performance. Ground truth
spectrum maps are generated using a calibrated propagation model that combines
distance-dependent path loss, frequency-selective attenuation, and obstacle-induced
occlusion losses computed via line-of-sight rasterization between transmitter and
receiver. While this simplified model omits explicit reflection and diffraction
kernels, it produces non-stationary field structures that capture the dominant
shadowing patterns in obstacle-dense environments and serves as a controlled
benchmark for evaluating sparse construction methods.  In the executed configuration, the UAV
surveys a fixed-altitude slice discretized into a $64\times64$ grid over four
monitored bands, the cognition input is a ten-channel tensor formed by four sparse
RSSI channels, four interpolated channels, the sampling mask, and the obstacle map,
and the extracted semantic embedding has dimension 1024. To evaluate
performance, ECDNet is compared against four representative baselines:
Kriging \cite{ref_1} (classical geostatistical interpolation), AE3D \cite{ref_2}
(purely data-driven autoencoder), U-Net \cite{ref_3} (semantic-assisted spatial
architecture), and KG-Net \cite{ref_4} (model-guided learning implemented via a
U-Net backbone with pyramid spatial pooling).  AE3D, U-Net, and KG-Net are
trained on the same 23,009-map split as ECDNet and selected according to validation
performance, and all construction methods share the same output grid, non-revisiting
spiral measurement locations, and RMSE definition.

Fig.~\ref{Fig.5} reports the training dynamics, where the upper panel shows cognition
pretraining and the lower panel shows the two policy-training stages. In the
cognition pretraining phase, purely data-driven methods typically struggle to find
optimal solutions within the high-dimensional parameter space and often lead to
slow convergence. In contrast, our method accelerates convergence by incorporating
the physics-informed transformation $\Phi_{phys}$, with the loss dropping rapidly
within the first 20 epochs. This indicates that the distance-aware
convolutions provide an inductive bias that constrains the solution space early in
optimization. For policy optimization, the Stage-2 refiner is updated by
PPO with $K=4$ parallel environments.  Here $K$ denotes the number of
environments used concurrently for rollout collection rather than the size of the
training set. The policy-training pool contains 540 scene instances from six scene
families, and Stage-2 refinement uses 90 instances of the obstacle-free
single-transmitter family (single-free) while the Stage-1 policy and the cognition
encoder remain frozen. The Stage-2 reward rises steadily while the
evaluation RMSE decreases consistently, indicating that the construction-aware
residual refinement effectively improves the exploration policy. All subsequent test-time evaluations are
strictly conducted on independent held-out instances to ensure fair comparison.

\begin{figure*}[!t]
\centering
\includegraphics[width=16.5cm,trim=5 5 5 5,clip]{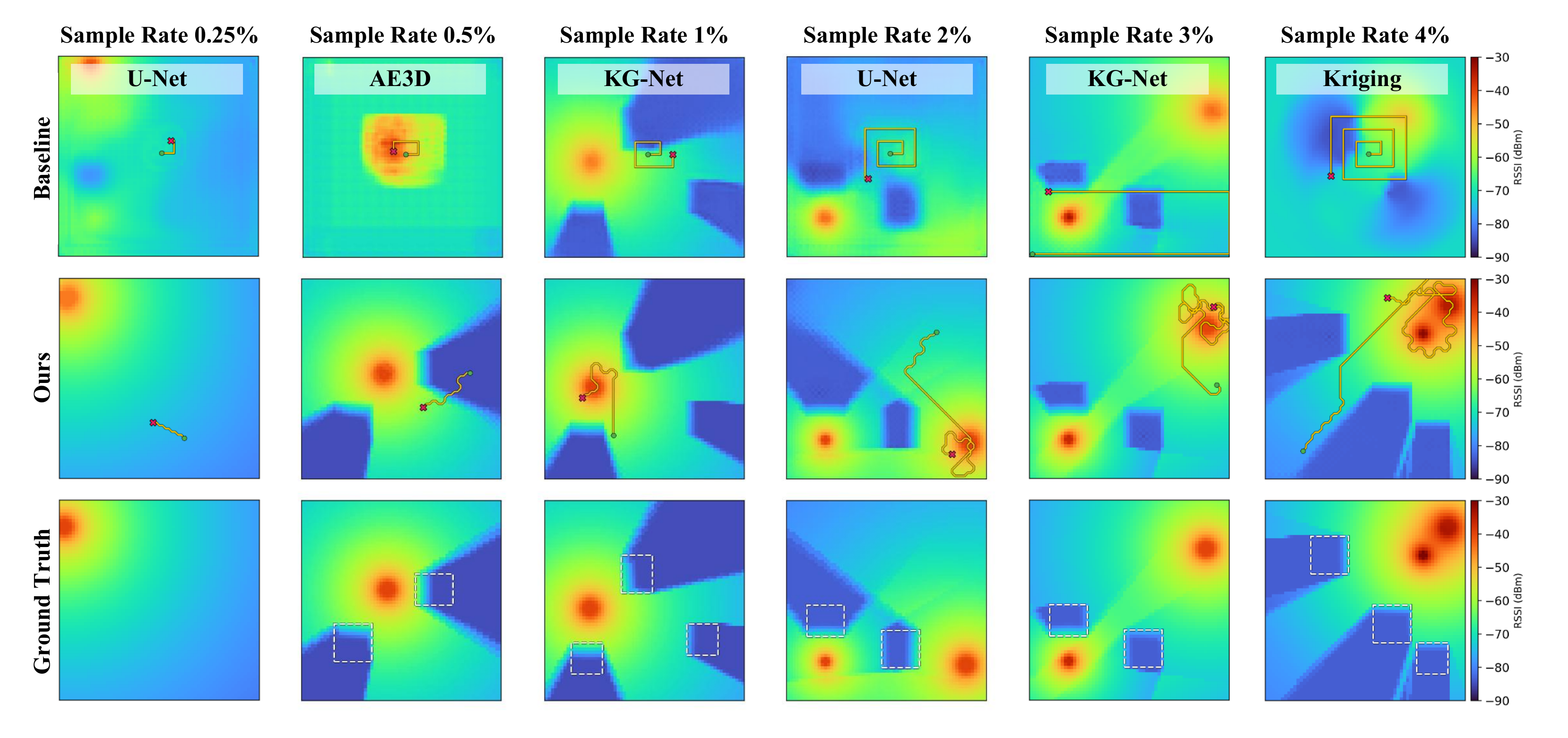}
\caption{Visual comparison of spectrum map construction quality and sampling trajectories across different methods at various sparsity levels. The panel headings are nominal rates, and the 0.25\% and 4\% panels correspond to 0.24\% and 4.00\% in Table~\ref{tab:spiral_metrics}. Baseline maps use the non-revisiting spiral samples, whereas Ours maps use the samples along the overlaid learned-policy trajectories.}

\label{Fig.7}
\end{figure*}

The spectrum construction performance is evaluated in Fig.~\ref{Fig.6} and
Table~\ref{tab:spiral_metrics} across different sampling rates. As illustrated in
Fig.~\ref{Fig.6}, spectrum construction faces marked degradation under extreme
sparsity conditions where the sampling rate falls below 2\%. Kriging and AE3D
exhibit high RMSE values in these regimes because they fail to capture the
non-stationary variations of shadowing without sufficient samples. In contrast,
ECDNet achieves the lowest RMSE in all three scenarios once the sampling rate
exceeds approximately 0.6\%, and its SSIM exceeds that of all baselines once the
sampling rate reaches approximately 1\%. Moreover, the RMSE of ECDNet continues to
decrease as measurements accumulate, whereas the baseline curves decrease slowly or
level off at higher error levels, demonstrating the superior sample efficiency of
ECDNet under sparse sampling.

Table~\ref{tab:spiral_metrics} further quantifies the construction accuracy at
discrete operating points,  where the sampling rate is $n/4096$ over the
$64\times64$ grid and $n$ counts the distinct sampled cells. As shown in
Table~\ref{tab:spiral_metrics}, ECDNet achieves the lowest RMSE and the highest SSIM
at all six operating points. Specifically, at 4.00\% sampling, ECDNet achieves an
RMSE of 2.787~dBm and an SSIM of 0.822, which reduces the RMSE by 47.8\% and
improves the SSIM by 0.149 compared with U-Net, the strongest baseline. 
Moreover, ECDNet obtains an RMSE of 4.042~dBm with only $n=41$ sampled cells
(1.00\%), which is still lower than the 5.335~dBm obtained by U-Net with $n=164$
sampled cells (4.00\%). This corresponds to a 75\% reduction in sampled locations at
this pair of operating points. These results demonstrate the superior sample
efficiency of ECDNet under extremely sparse sampling.

\begin{table*}[t]
\centering
\caption{Quantitative performance comparison under the non-revisiting spiral sampling pattern.}

\label{tab:spiral_metrics}
\begin{tabular}{lcccccccccccc}
\toprule
\multirow{2}{*}{Method} & \multicolumn{2}{c}{0.24\% (n=10)} & \multicolumn{2}{c}{1.00\% (n=41)} & \multicolumn{2}{c}{2.00\% (n=82)} & \multicolumn{2}{c}{3.00\% (n=123)} & \multicolumn{2}{c}{4.00\% (n=164)} & \multicolumn{2}{c}{4.88\% (n=200)} \\
\cmidrule(lr){2-3}\cmidrule(lr){4-5}\cmidrule(lr){6-7}\cmidrule(lr){8-9}\cmidrule(lr){10-11}\cmidrule(lr){12-13}
& RMSE & SSIM & RMSE & SSIM & RMSE & SSIM & RMSE & SSIM & RMSE & SSIM & RMSE & SSIM \\
\midrule
AE3D \cite{ref_2} & 12.058 & 0.454 & 11.670 & 0.454 & 11.296 & 0.457 & 10.896 & 0.460 & 10.525 & 0.475 & 10.093 & 0.482 \\
Kriging \cite{ref_1} & 13.096 & 0.605 & 12.468 & 0.585 & 12.048 & 0.565 & 11.827 & 0.552 & 12.117 & 0.529 & 12.164 & 0.512 \\
U-Net \cite{ref_3} & 11.005 & 0.483 & 8.078 & 0.589 & 6.335 & 0.646 & 5.781 & 0.665 & 5.335 & 0.673 & 4.924 & 0.684 \\
KG-Net \cite{ref_4} & 11.428 & 0.508 & 10.118 & 0.547 & 9.114 & 0.570 & 8.527 & 0.582 & 8.098 & 0.588 & 7.769 & 0.593 \\
\textbf{Ours} & \textbf{6.689} & \textbf{0.644} & \textbf{4.042} & \textbf{0.780} & \textbf{3.155} & \textbf{0.810} & \textbf{2.847} & \textbf{0.820} & \textbf{2.787} & \textbf{0.822} & \textbf{2.746} & \textbf{0.823} \\
\bottomrule
\end{tabular}
\end{table*}

\begin{table*}[t]
\centering

\caption{Equal-budget active-exploration controls on the 60-episode validation protocol. All controls use 205 steps and the same legal-action mask.}

\label{tab:active_controls}
\begin{tabular}{lccc}
\toprule
Controller & Evaluation seed & Minimum RMSE (dBm) & Final RMSE (dBm) \\
\midrule
Learned policy & 7471 & 1.622 & 2.877 \\
Random legal action & 7531 & 2.392 & 4.411 \\
Uniform legal-action cycle & 7551 & 2.539 & 4.046 \\
\bottomrule
\end{tabular}
\end{table*}

\begin{table*}[t]
\centering

\caption{Family-level performance of the selected checkpoint. The
cognition encoder and the Stage-1 policy are frozen, so that only the Stage-2 refiner
heads differ across candidate checkpoints. Validation uses 10 episodes per family with
base seed 31415, and the independent locked test split uses 20 episodes per family with
base seed 271828. Aggregate values are equal-weight means across the six families.}

\label{tab:family_results}
\begin{tabular}{lrrrr}
\toprule
Scene family & Validation min & Validation final & Locked min & Locked final \\
\midrule
Double-free, fixed weak & 0.838 & 1.226 & 0.892 & 1.113 \\
Double-obstacle layout A & 1.679 & 1.992 & 1.543 & 2.363 \\
Double-obstacle layout B & 0.964 & 1.227 & 1.643 & 2.627 \\
Single-free full grid & 2.712 & 6.590 & 4.239 & 8.160 \\
Single obstacle, 3 rectangles & 1.620 & 2.088 & 1.485 & 1.912 \\
Single obstacle, 2 rectangles & 1.143 & 1.492 & 1.300 & 1.543 \\
\midrule
Aggregate & 1.492 & 2.436 & 1.850 & 2.953 \\
\bottomrule
\end{tabular}
\end{table*}

 Fig.~\ref{Fig.7} shows representative spectrum-map estimates at sampling
rates from 0.24\% to 4.00\%. As shown in the top row of Fig.~\ref{Fig.7},
baseline methods including U-Net and Kriging tend to produce over-smoothed results or
blocky artifacts due to the ill-posed nature of sparse construction. For instance,
the AE3D model at the 0.5\% sampling rate fails to capture the distinct signal
propagation direction while Kriging at the 4\% rate exhibits characteristic smoothing
artifacts around sampling points.  In contrast, as shown in the middle row,
ECDNet recovers the dominant signal sources and the obstacle-induced shadow
boundaries even at low sampling rates, and its constructions are highly consistent
with the ground truth. Unlike the spiral-sampled baselines, the ECDNet maps in
Fig.~\ref{Fig.7} are constructed along the overlaid learned-policy trajectories,
which actively visit shadow regions and the vicinity of signal sources, and thus
reflect both the propagation-aware encoder and construction-aware sampling, whereas
Table~\ref{tab:spiral_metrics} compares construction at identical sampling
locations.

 Table~\ref{tab:active_controls} compares the learned policy with random and
uniform legal-action controls under a common budget of 205 steps, which covers at
most $205/4096=5.005\%$ of the grid because revisits are permitted. All controls
share the same history-based legal-action mask, under which an action is illegal if
it would leave the grid, enter an obstacle, or exceed the revisit threshold within
the recent-visit window. The minimum RMSE is the lowest map RMSE reached within an
episode, and the final RMSE is the value at its last step. As shown in
Table~\ref{tab:active_controls}, the learned policy achieves the lowest
minimum and final RMSE of 1.622 and 2.877~dBm, which are 32.2\% and 34.8\% lower than
those of the random legal-action control and 36.1\% and 28.9\% lower than those of
the uniform legal-action cycle, respectively. These results demonstrate that the
construction-aware policy collects more informative measurements than non-learned
exploration under the same action budget.

 Table~\ref{tab:family_results} reports the family-level performance of the
selected checkpoint. Although Stage-2 refinement uses only single-free instances,
the checkpoint is selected on the six-family validation aggregate of 60 episodes and
then evaluated once on an independent locked test split of 120 episodes, which is
never used for model selection. As shown in Table~\ref{tab:family_results}, the selected checkpoint achieves
aggregate minimum RMSE values of 1.492 and 1.850~dBm on the validation and locked-test
splits, and five of the six families remain below 1.65~dBm on the locked test split.
Table~\ref{tab:active_controls} evaluates the same checkpoint with a different seed
and mask setting, $(8,0.50)$ instead of $(4,0.75)$, and each comparison is therefore
made within its own protocol group. These results demonstrate that the learned
policy remains effective across different numbers of active transmitters and obstacle
layouts, while the single-free family is the most demanding because the full free
grid provides no obstacle boundary as a spatial anchor. Moreover, every family
reaches its minimum RMSE before the 205-step budget is exhausted, indicating that
the best maps are obtained within the step budget.

\begin{figure*}[!t]
\centering
\includegraphics[width=16cm,trim=1 1 1 1,clip]{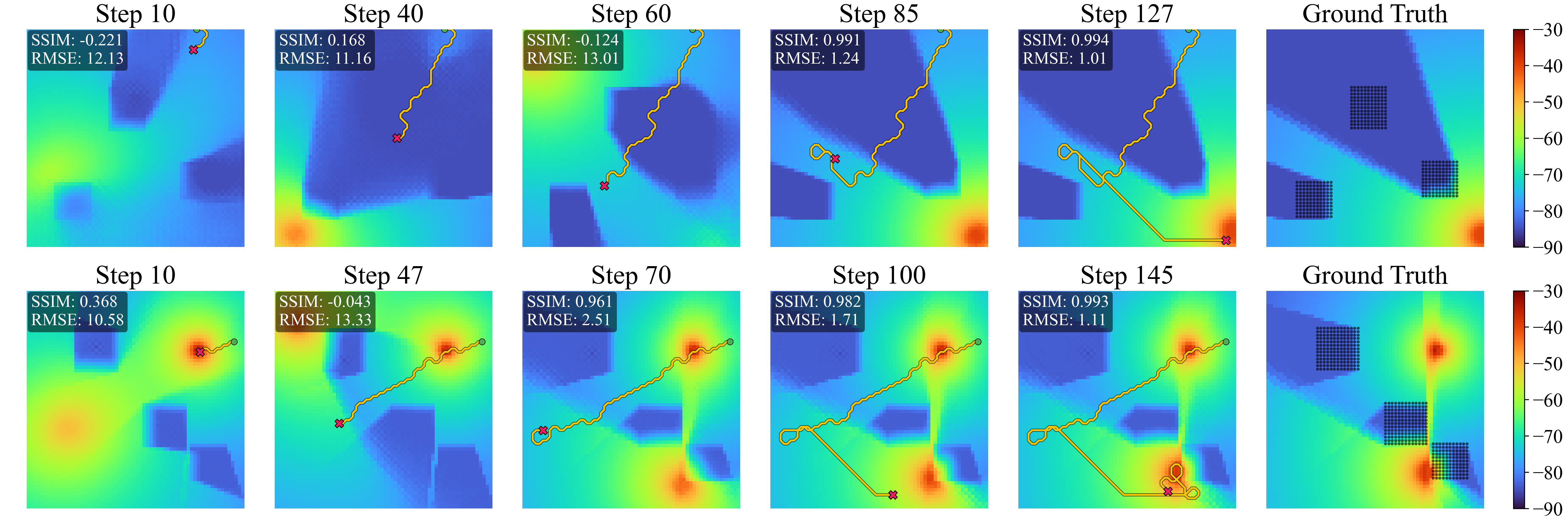}
\caption{Progressive evolution of the spectrum map and UAV active exploration trajectory during online missions in representative cases.}

\label{Fig.8}
\end{figure*}

\begin{figure}[t]
\centering
\includegraphics[width=8cm,trim=5 5 5 5,clip]{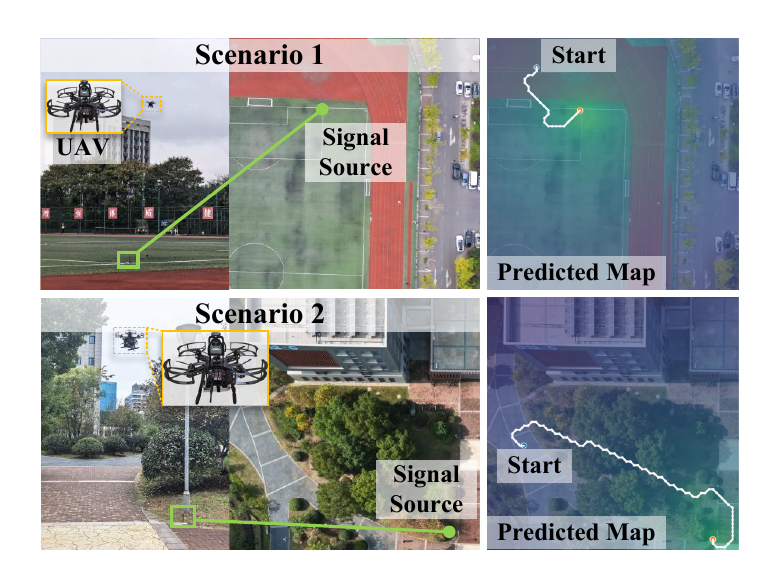}
\caption{Prototype-level field demonstrations in an open campus setting (Scenario~1)
and a forest setting with vegetation and nearby obstacles (Scenario~2).}

\label{Fig.9}
\end{figure}

To further reveal the internal coupling mechanism of the PCE loop, Fig.~\ref{Fig.8}
illustrates the progressive evolution of the spectrum map and the corresponding UAV
trajectory during active exploration missions. In the early exploration phase shown
in Step 10 through Step 40, the UAV exhibits wide-range movement behavior to cover
the survey area while the constructed map remains blurry with high error rates.
 As the UAV moves along the obstacle boundaries, the RMSE of the upper episode drops
sharply from 13.01~dBm at Step 60 to 1.24~dBm at Step 85. In the lower episode,
the UAV makes local movements between the two signal sources from Steps 100 to 145,
and the final map reaches an RMSE of 1.11~dBm and an SSIM of 0.993.

Finally, the field deployment results in real-world campus and forest environments
are presented in Fig.~\ref{Fig.9}. Real-world deployment introduces the challenge
of domain shift where unknown interference and hardware noise can degrade model
performance in practical settings. As shown in Scenario 1 in the top row of
Fig.~\ref{Fig.9}, the UAV operates in an open campus environment with a
line-of-sight signal source placed on the sports field.  The predicted map
exhibits a high-RSSI region consistent with the known location of the signal
source. The
bottom row of Fig.~\ref{Fig.9} depicts Scenario 2, which presents a more
challenging environment featuring dense vegetation and nearby obstacles. This
setting is expected to introduce stronger shadowing and multipath effects that
obscure the signal.  Nevertheless, the UAV successfully navigates through
the obstacle-rich forest setting while continuously updating the spectrum map. In
both settings, the complete chain from sensing to action selection runs onboard the
Jetson Orin NX and produces a $64\times64$ spectrum-map output, demonstrating the
prototype-level feasibility of ECDNet for closed-loop spectrum map construction on
SWaP-constrained UAV platforms.

\section{Conclusion}

In this paper, an embodied intelligence framework for spectrum map construction was proposed to achieve precise spectrum construction through closed-loop PCE coupling and embodied exploration.
This framework was instantiated as the ECDNet, which effectively integrated the PCE loop through distance-aware convolutions, semantic bridge extraction, and cognition-guided
decision modules.
 Furthermore, a practical UAV-based spectrum construction platform was developed for prototype-level evaluation in two field settings.
The experimental results show that ECDNet achieves 2.787~dBm RMSE at 4.00\% sampling under the non-revisiting spiral protocol and, with only 1.00\% sampling, attains a lower RMSE than U-Net at 4.00\%, corresponding to 75\% fewer sampled locations at this pair of operating points.
Moreover, the field tests provide prototype-level evidence of onboard closed-loop operation.
These findings support EI-based spectrum map construction within the six evaluated simulation scene families and the two field settings, and the framework offers a basis for interference mitigation and spectrum-aware trajectory planning in LAWNs.

\vfill
\end{document}